\documentclass[a4paper,fleqn,usenatbib]{pasa}

\title[The Taipan Galaxy Survey]{The Taipan Galaxy Survey: \\
Scientific Goals and Observing Strategy}

\author[E. da Cunha et al.]{Elisabete da Cunha,$^{1\thanks{E-mail: elisabete.dacunha@anu.edu.au}}$
Andrew M. Hopkins,$^{2}$
Matthew Colless,$^{1}$
Edward N. Taylor,$^{3}$
Chris Blake,$^{3}$
Cullan Howlett,$^{4,5}$
Christina Magoulas,$^{1,6}$
John R. Lucey,$^{19}$
Claudia Lagos,$^{4}$
Kyler Kuehn,$^{2}$
Yjan Gordon,$^{7}$
Dilyar Barat,$^{1}$
Fuyan Bian,$^{1}$
Christian Wolf,$^{1}$
Michael J. Cowley,$^{8,9,2}$
Marc White,$^{1}$
Ixandra Achitouv,$^{3,5}$
Maciej Bilicki,$^{10,11}$
Joss Bland-Hawthorn,$^{12}$
Krzysztof Bolejko,$^{12}$
Michael J. I. Brown,$^{13}$
Rebecca Brown,$^{2}$
Julia Bryant,$^{12,2,5}$
Scott Croom,$^{12}$
Tamara M. Davis,$^{14}$
Simon P. Driver,$^{4}$
Miroslav D. Filipovic,$^{21}$
Samuel R. Hinton,$^{14}$
Melanie Johnston-Hollitt,$^{15,16}$
D. Heath Jones,$^{17}$
B\"arbel Koribalski,$^{18}$
Dane Kleiner,$^{18}$
Jon Lawrence,$^{2}$ 
Nuria Lorente,$^{2}$
Jeremy Mould,$^{3}$
Matt S. Owers,$^{8,2}$
Kevin Pimbblet,$^{7}$
C. G. Tinney,$^{20}$
Nicholas F. H. Tothill,$^{21}$
Fred Watson$^{2}$ 
\\
\affil{$^1$Research School of Astronomy and Astrophysics, Australian National University, Canberra, ACT 2611, Australia}
\affil{$^2$Australian Astronomical Observatory, 105 Delhi Rd., North Ryde, NSW 2113, Australia}
\affil{$^3$Centre for Astrophysics \& Supercomputing, Swinburne University of Technology, P.O.Box 218, Hawthorn, VIC 3122, Australia}
\affil{$^4$International Centre for Radio Astronomy Research, University of Western Australia, Crawley, WA 6009, Australia}
\affil{$^5$ARC Centre of Excellence for All-sky Astrophysics (CAASTRO), 44 Rosehill St, Redfern, NSW 2016, Australia}
\affil{$^6$Department of Astronomy, University of Cape Town, Private Bag X3, Rondebosch 7701, South Africa}
\affil{$^7$E.A. Milne Centre for Astrophysics, University of Hull, Cottingham Road, Kingston upon Hull HU6 7RX, United Kingdom}
\affil{$^8$Department of Physics and Astronomy, Macquarie University, NSW 2109, Australia}
\affil{$^9$Research Centre for Astronomy, Astrophysics \& Astrophotonics, Macquarie University, Sydney, NSW 2109, Australia}
\affil{$^{10}$Leiden Observatory, Leiden University, P.O. Box 9513, NL-2300 RA Leiden, The Netherlands}
\affil{$^{11}$National Centre for Nuclear Research, Astrophysics Division, P.O. Box 447, PL-90-950 Lodz, Poland}
\affil{$^{12}$Sydney Institute for Astronomy (SIfA), School of Physics, The University of Sydney, NSW 2006, Australia}
\affil{$^{13}$Monash Centre for Astrophysics, Monash University, Clayton, Victoria 3800, Australia}
\affil{$^{14}$School of Mathematics and Physics, The University of Queensland, Brisbane, QLD 4072, Australia}
\affil{$^{15}$School of Chemical \& Physical Sciences, Victoria University of Wellington, PO Box 600, Wellington 6140, New Zealand}
\affil{$^{16}$Peripety Scientific Ltd., PO Box 11355 Manners Street, Wellington, 6142, New Zealand}
\affil{$^{17}$ English Language and Foundation Studies Centre, University of Newcastle, Callaghan, NSW 2308, Australia}
\affil{$^{18}$CSIRO Astronomy and Space Science, Australia Telescope National Facility, PO Box 76, Epping, NSW 1710, Australia}
\affil{$^{19}$Centre for Extragalactic Astronomy, University of Durham, Durham DH1 3LE, United Kingdom}
\affil{$^{20}$Exoplanetary Science at UNSW School of Physics, University of New South Wales, Sydney, NSW 2052, Australia}
\affil{$^{21}$School of Computing, Engineering and Mathematics, Western Sydney University, Locked Bag 1797, Penrith, NSW 2751, Australia}
}

\jid{PASA}
\doi{10.1017/pas.\the\year.xxx}
\jyear{\the\year}

\usepackage[authoryear]{natbib}
\bibpunct{(}{)}{;}{a}{}{,}
\usepackage{aas_macros,enumitem}
\usepackage{hyperref} 
\hypersetup{colorlinks,citecolor=blue,linkcolor=blue,urlcolor=blue}

\usepackage{enumitem,rotating,lscape}

\usepackage{graphicx}	

\usepackage{amsmath}	
\usepackage{amssymb}	
\usepackage[usenames, dvipsnames]{color}
\usepackage{newtxtext,newtxmath}
\usepackage[T1]{fontenc}
\usepackage{ae,aecompl}
\usepackage{times}

\newcommand{\ho}{\hbox{$H_0$}}
\newcommand{\msun}{\hbox{$\mathrm{M}_{\odot}$}}

\begin{document}

\begin{abstract}
The Taipan galaxy survey (hereafter simply `Taipan') is a multi-object spectroscopic survey starting in 2017 that will cover 2$\pi$ steradians over the southern sky ($\delta\lesssim10\deg$, $|b|\gtrsim 10\deg$), and obtain optical spectra for about two million galaxies out to $z < 0.4$.
Taipan will use the newly-refurbished 1.2-metre UK Schmidt Telescope at Siding Spring Observatory with the new TAIPAN instrument, which includes an innovative `Starbugs' positioning system capable of rapidly and simultaneously deploying up to 150 spectroscopic fibres (and up to 300 with a proposed upgrade) over the 6-degree diameter focal plane, and a purpose-built spectrograph operating in the range from $370$ to $870\,$nm with resolving power $R\gtrsim2000$.
The main scientific goals of Taipan are:
(i) to measure the distance scale of the Universe (primarily governed by the local expansion rate, $H_0$) to 1\% precision, and the growth rate of structure to 5\%;
(ii) to make the most extensive map yet constructed of the total mass distribution and motions in the local
Universe, using peculiar velocities based on improved Fundamental Plane distances, which will enable sensitive tests of gravitational physics; and
(iii) to deliver a legacy sample of low-redshift galaxies as a unique laboratory for studying galaxy evolution as a function of dark matter halo and stellar mass and environment.
The final survey, which will be completed within five years, will consist of a complete magnitude-limited sample ($i\le17$) of about $1.2\times10^6$ galaxies, supplemented by an extension to higher redshifts and fainter magnitudes ($i\le18.1$) of a luminous red galaxy sample of about $0.8\times10^6$ galaxies.
Observations and data processing will be carried out remotely and in a fully-automated way, using a purpose-built  automated `virtual observer' software and an automated data reduction pipeline.
The Taipan survey is deliberately designed to maximise its legacy value, by complementing and enhancing current and planned surveys of the southern sky at
wavelengths from the optical to the radio; it will become the primary redshift and optical spectroscopic reference catalogue for the local extragalactic Universe in the southern sky for the coming decade.
\end{abstract}

\begin{keywords}
surveys -- techniques: spectroscopic -- cosmology: observations -- galaxies: distances and redshifts
\end{keywords}

\maketitle


\section{Introduction}

Large extragalactic spectroscopic surveys carried out in the last few decades have enormously improved our understanding of the content and evolution of the Universe. These surveys include the 2-degree Field Galaxy Redshift Survey (2dFGRS; \citealt{Colless2001}), the Sloan Digital Sky Survey (SDSS; \citealt{York2000,Eisenstein2001,Abazajian2009,Dawson2016}), the 6-degree Field Galaxy Survey (6dFGS; \citealt{Jones2004, Jones2009}), the Galaxy And Mass Assembly survey (GAMA; \citealt{Driver2011,Hopkins2013,Liske2015}), the WiggleZ Dark Energy Survey \citep{Drinkwater2010}, and the Baryon Oscillation Spectroscopic Survey (BOSS; \citealt{Dawson2013,Reid2016}). Using these surveys, we have started making detailed maps of the baryonic and dark matter distribution and bulk motions in the local Universe (e.g.\,\citealt{Springob2014,Scrimgeour2016}), constraining cosmological models with increasing precision (e.g.\,\citealt{Beutler2011,Blake2011b,Anderson2012,Johnson2014,Alam2016}), and obtaining a census of the properties of present-day galaxies (e.g.\,\citealt{Kauffmann2003,Blanton2009,Baldry2012,Liske2015,Lange2016,Moffett2016}).
The value of these major spectroscopic programmes comes not only from their primary scientific drivers,
but also from the legacy science they facilitate by making large optical datasets publicly available which enables novel and unforeseen science, especially in conjunction with datasets at other wavelengths (e.g.\,\citealt{Driver2016}).

Here we describe the Taipan galaxy survey. This new southern hemisphere spectroscopic survey will complement and enhance the results from earlier large-scale survey projects. Specifically, Taipan will extend beyond the depth of  6dFGS, and increase by an order of magnitude the number of galaxies
with optical spectra measured over the whole southern hemisphere, enabling major programmes in both cosmology and galaxy evolution in the nearby Universe.
The survey strategy is designed to optimally achieve three main goals:
\begin{enumerate}[label=(\roman*),topsep=0pt]
\item To measure the present-day distance-scale of the Universe (which is principally governed by the Hubble parameter \ho) with 1\% precision, and the
growth rate of structure to 5\%. This will represent an improvement by a factor of four over current low-redshift distance constraints from baryon acoustic oscillations (BAOs) \citep{Beutler2011,Ross2015}, and by a factor of two over the best existing standard-candle determinations \citep{Riess2016}.
\item To make the most extensive map yet constructed of the motions of matter (as traced by galaxies) in the local
Universe, using peculiar velocities for a sample more than five times larger than 6dFGS (the largest homogeneous peculiar velocity survey to date), combined with improved Fundamental Plane constraints.
\item To determine in detail, in the redshift and magnitude ranges probed, the baryon lifecycle, and the role of halo mass, stellar mass, interactions, and large-scale environment in the evolution of galaxies.
\end{enumerate}
Extending the depth of 6dFGS with Taipan (and maximising the volume probed), leads directly to the opportunity for improvements to the main scientific results arising from 6dFGS. Specifically, this includes using the BAO technique for measuring the distance-scale of the low-redshift Universe (e.g.\,\citealt{Beutler2011}), and using galaxy peculiar velocities to map gravitationally-induced motions (e.g.\,\citealt{Springob2014}). With the precision enabled by the scale of the Taipan survey, we will make stringent tests of cosmology by comparing to predictions from the cosmological Lambda Cold Dark Matter ($\Lambda$CDM) model and from the theory of General Relativity.

Furthermore, with the ability to provide high-completeness ($>98\%$) sampling of the galaxy population at low redshift, we will explore the role of interactions and the environment in galaxy evolution. This is enabled by the multiple-pass nature of the
Taipan survey, ensuring that high-density regions of galaxy groups and clusters are well-sampled. Taipan will be combined with the upcoming wide-area neutral
hydrogen (HI) measurements from the Wide-field ASKAP L-band Legacy All-sky Blind surveY (WALLABY; \citealt{Koribalski2012}) with the Australian Square Kilometre Array Pathfinder (ASKAP; \citealt{Johnston2008}), which will probe a similar redshift range. This combination will lead to a comprehensive census of baryons in the low-redshift Universe, and the opportunity to follow the flow of baryons from HI to stellar mass through star formation processes, and to quantify how these processes are influenced by galaxy mass, close interactions, and the large-scale environment.

There is a significant effort worldwide to expand the photometric survey coverage of the southern hemisphere, including radio surveys with the Murchison Widefield Array (MWA; \citealt{Tingay2013}), the Square Kilometre Array pathfinder telescopes in Australia (ASKAP; \citealt{Johnston2008}) and South Africa (MeerKAT; \citealt{Jones2009}), infrared surveys with the Visible and Infrared Survey Telescope for Astronomy (VISTA; e.g.\,\citealt{McMahon2013}), and optical surveys with the Panoramic Survey Telescope and Rapid Response System (Pan-STARRS; \citealt{Kaiser2010,Chambers2016}), the VLT Survey Telescope (VST; \citealt{Kuijken2011}), and SkyMapper \citep{Keller2007}.
The need for hemispheric coverage with optical spectroscopy to maximise the scientific return from all these programmes is clear.
In addition to the main scientific motivations for the Taipan galaxy survey described above, the legacy value of the project will be substantial. Taipan will complement these and other Southern surveys (e.g.\,Hector; \citealt{BlandHawthorn2015}), and it will provide the primary redshift and optical spectroscopic reference for the southern hemisphere for the next decade.
Imaging and spectroscopic mapping of the southern sky will be continued in the future by the Large Synoptic Survey Telescope (LSST; \citealt{Tyson2002}), the {\em Euclid} satellite \citep{Racca2016}, the Square Kilometre Array (SKA; e.g.\,\citealt{Dewdney2009}), Cosmic Microwave Background (CMB) Stage 4 Experiment \citep{Abazajian2016}, the 4-metre Multi-Object Spectrograph Telescope (4MOST; \citealt{deJong2012}), and the eROSITA space telescope in the X-rays \citep{Merloni2012}.

Taipan will be conducted with the newly-refurbished 1.2\,m UK Schmidt Telescope (UKST) at Siding Spring Observatory, Australia. It will use the new `Starbugs' technology developed at the Australian Astronomical Observatory (AAO), which allows the rapid and simultaneous deployment of 150 spectroscopic fibres (and up to 300 with a proposed upgrade) over the 6-degree focal plane of the UKST.
The use of optical fibres to exploit the wide field of the UKST was first proposed in a memorandum of July 1st, 1982 \citep{Dawe1982}. Thirty-five years later, the technology proposed in that note has come to fruition with the TAIPAN instrument.\footnote{We note that `TAIPAN' refers to the instrument system on the UKST, while `Taipan' or `Taipan survey' refers to the galaxy survey.} Four generations of multi-fibre spectroscopy systems have preceded it on the UKST: FLAIR (1985), PANACHE (1988), FLAIR II (1992) and 6dF (2001) (see \citealt{Watson2011}, and references therein). The prototype FLAIR was the first multi-fibre instrument on any telescope to feed a stationary spectrograph, and the first truly wide-field multi-object spectroscopy system. Its successors generated a wide and varied body of data, most notably the 6dFGS and Radial Velocity Experiment (RAVE; e.g.\,\citealt{Steinmetz2006}) surveys.

The Starbugs technology on TAIPAN dramatically increases the survey speed and efficiency compared to previous large-area southern surveys, and it will allow us within five years to obtain about two million galaxy spectra covering the whole southern hemisphere to an optical magnitude limit approaching that of SDSS. Thus, Taipan will be the most comprehensive spectroscopic survey of the southern sky performed to date.

Taipan will be executed using a two-phase approach, driven by the availability of input photometric catalogues for target selection, as well as a planned upgrade from 150 to 300 fibres during the course of the survey. {\em Taipan Phase~1} will run from late-2017 to the end of 2018, and a second {\em Taipan Final} phase will run from the start of 2019 to the end of main survey operations. This strategy will allow us to maximise the early scientific return of Taipan, with the Taipan Phase~1 sample being contained in the Taipan Final sample.

In this paper, we introduce the Taipan galaxy survey and its goals, and describe the data acquisition and processing strategy devised to achieve those goals. This paper is organised as follows. In Section~\ref{instrument}, we describe the purpose-built TAIPAN instrument used to carry out our observations on the UKST. In Section~\ref{goals}, we describe the main scientific goals of the Taipan galaxy survey, and in Section~\ref{strategy} we outline the survey strategy, including target selection, observing and data processing strategy, and plans for data archiving and dissemination. A summary and our conclusions are presented in Section~\ref{conclusion}.

Throughout the paper, we use AB magnitudes, and a $\Lambda$CDM cosmology with $\ho=100h$~km~s$^{-1}$~Mpc$^{-1}$, $h=0.7$, $\Omega_\Lambda=0.7$, and $\Omega_m=0.3$, unless otherwise stated.

\section{The TAIPAN instrument}
\label{instrument}

\begin{figure}
 \includegraphics[trim={-1cm 1.5cm 3.5cm 0cm},width=0.65\columnwidth]{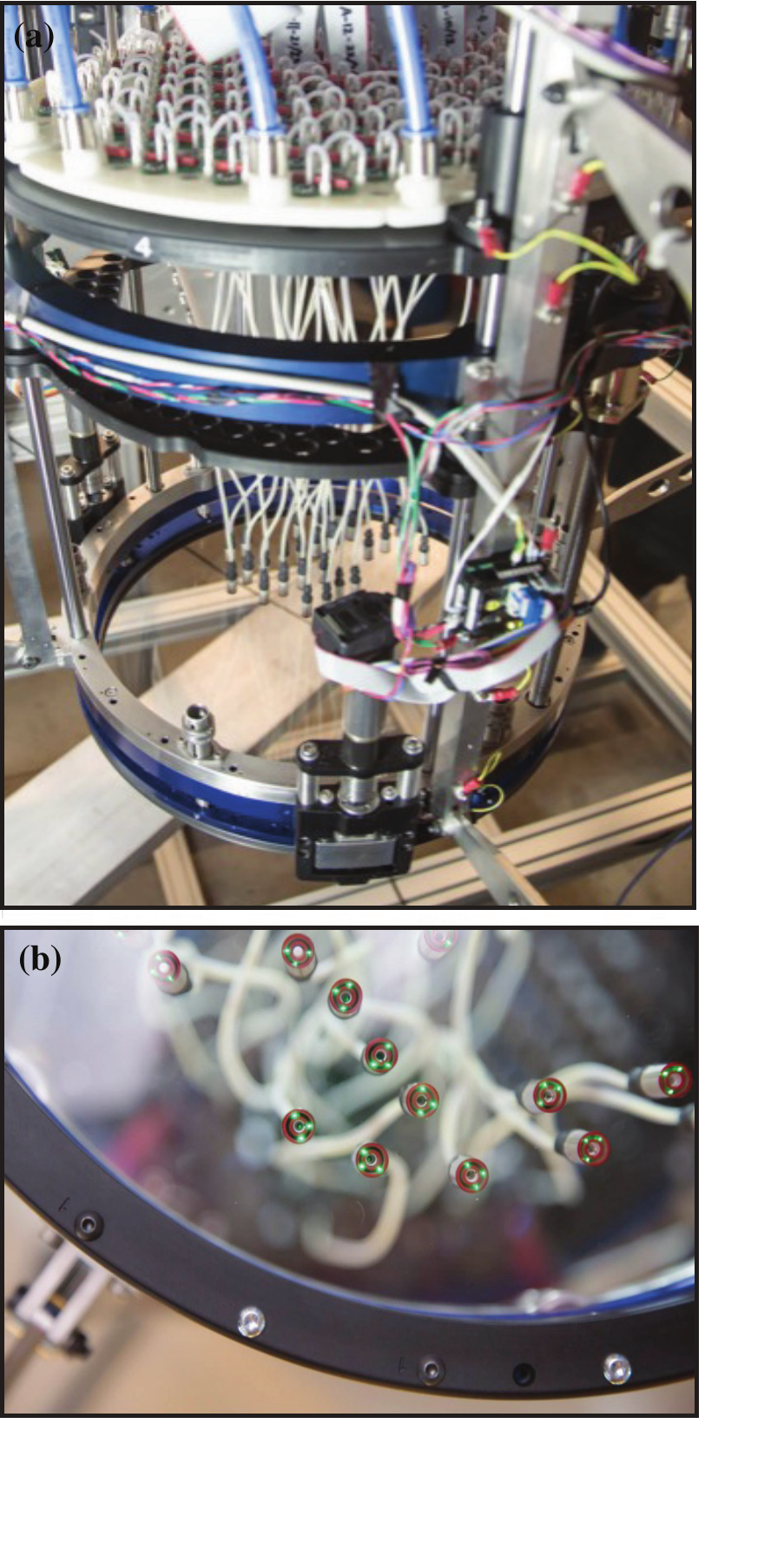}
 \caption{The TAIPAN fibre positioner at the AAO. {\bf (a)} Top view, showing 24 Starbugs installed on the glass field plate that sits at the focal surface of the UKST. The top of the image shows the complex vacuum and high-voltage support systems required for operation. {\bf (b)} Underside view, showing some of the 24 Starbugs installed on the glass field plate.}
  \label{taipan_picture}
\end{figure}

\begin{table}
\centering
 \caption{TAIPAN instrument specifications.}
 \label{tab:instrument}
 \begin{tabular}{ll}
  \hline
Field of view diameter & $6\,$degrees \\
Number of fibres       & 150 \\
                        & (300 planned from 2019 onwards) \\
 Fibre diameter         & $3.3$\,arcsec \\
 Wavelength range       & $370$ -- $870\,$nm\\
 Resolving power  & $1960 \equiv 65\,$km\,s$^{-1}$ (blue); \\
   ($\lambda/\Delta\lambda$)                                     & $2740 \equiv 46\,$km\,s$^{-1}$ (red) \\  \hline
 \end{tabular}
\end{table}

\begin{figure}
 \includegraphics[trim={2cm 0cm 0.5cm 0cm},width=\columnwidth]{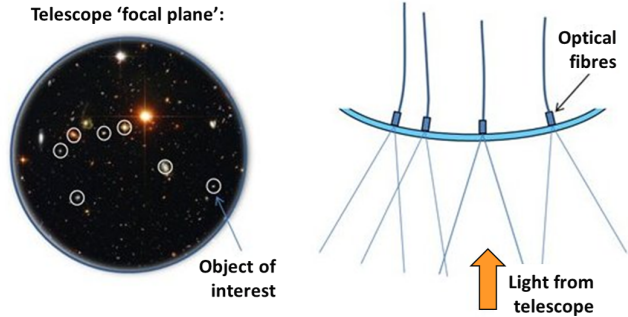}
  \caption{Schematic representation of the TAIPAN focal plane. The background image shows target galaxies as `objects of interest', while Starbugs are the depicted by the white open circles. Starbugs can move independently to put a spectroscopic fibre on any object of interest in the 6-degree diameter field of view. The right-hand side shows a side view.}
   \label{taipan_field}
\end{figure}

\begin{figure}
 \includegraphics[trim={0.75cm 1cm 0.75cm 0cm},width=\columnwidth]{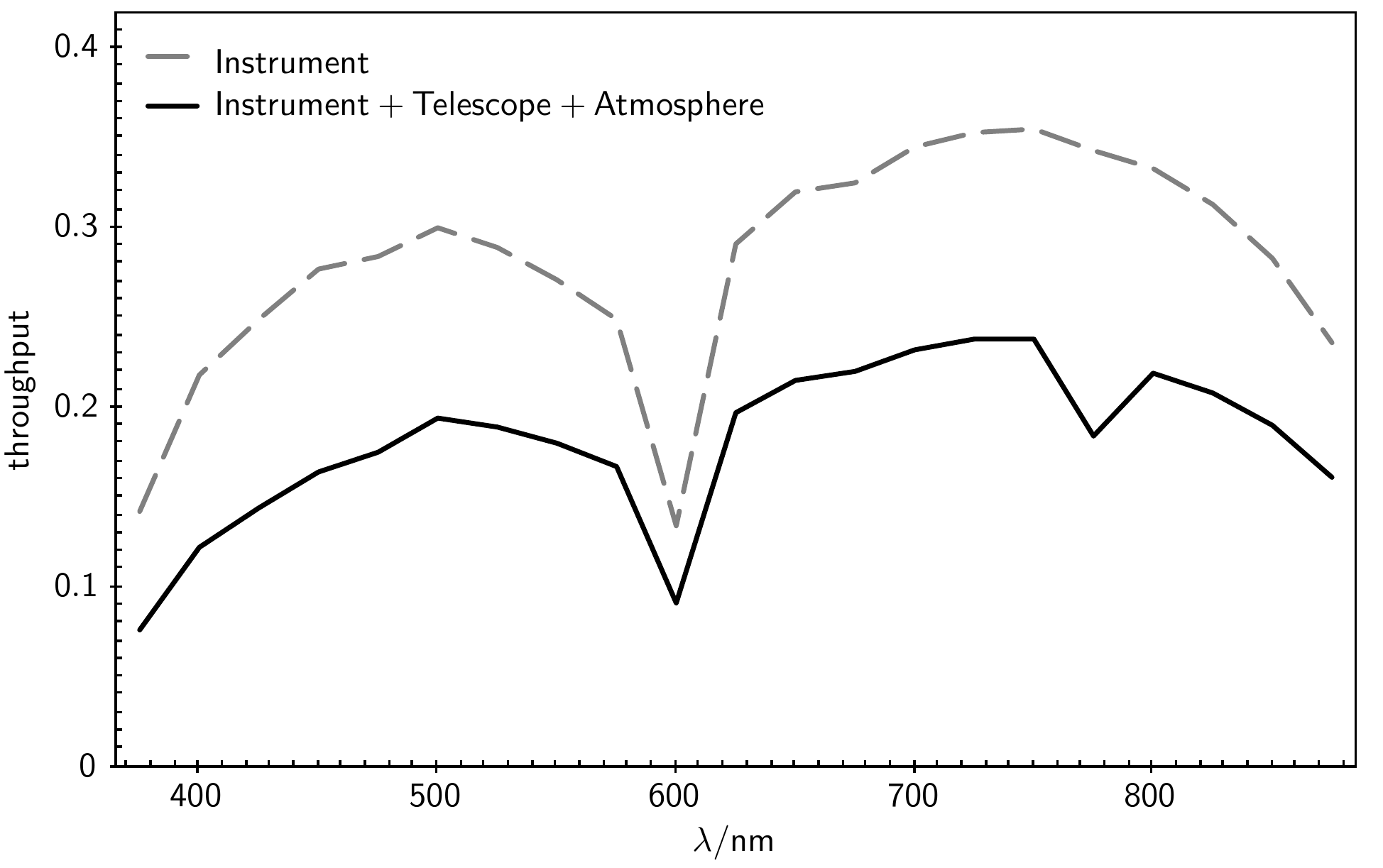}
  \vspace{0.05cm}
 \caption{Anticipated throughput of the TAIPAN instrument (dashed grey line), and total throughput of the whole system (i.e.~instrument+telescope+atmosphere; see also \protect\citealt{Kuehn2014}.)}
 \label{throughput}
\end{figure}

The TAIPAN instrument consists of a large multiplexed robotic fibre positioner operating over the 6-degree diameter field of view of the upgraded UK Schmidt Telescope, along with a dedicated spectrograph. The instrument specifications are summarised in Table~\ref{tab:instrument}. The fibre positioner (Fig.~\ref{taipan_picture}) is based on the Starbug technology \citep{Lorente2015} developed at the AAO, which enables the parallel repositioning of hundreds of optical fibres. TAIPAN will start with 150 science fibres, with a planned upgrade to 300 fibres to be available from 2019. Serial positioning robots, e.g., those used by the 2dF \citep{Lewis2002} or 6dF \citep{Jones2004}, accomplish field reconfigurations in tens of minutes to an hour -- the parallel positioning capability of Starbugs allows for field reconfiguration in less than five minutes.

During reconfiguration and observing, the Starbugs are held by a vacuum onto a glass plate curved to follow the focal surface of the telescope (Figs.~\ref{taipan_picture} and \ref{taipan_field}). Starbugs move by means of coaxial piezoceramic tubes to which high-voltage waveforms are applied. The resulting deformation of the piezoceramic `walks' the Starbugs across the glass plate. In addition to a centrally-located science fibre payload, each Starbug includes a trio of back-illuminated fibres that are viewed from beneath by a metrology camera to deliver  accurate Starbug positioning (Fig.~\ref{taipan_picture}). At the plate scale of the UKST, position uncertainty must be better than $5\,$microns to ensure the science fibres are positioned on the selected targets. Once the metrology system determines that the Starbugs are positioned with sufficient accuracy, light from the selected targets enters the central science fibre and travels $\sim20\,$m to the TAIPAN spectrograph. Within the spectrograph, the light from each fibre is split into blue ($370-592\,$nm) and red ($580-870\,$nm) components by a dichroic, and sent to two separate cameras, each with a 2k$\times$2k e2V CCD (\citealt{Kuehn2014}, see Fig.~\ref{throughput}).
While the spectroscopic fibres are only $3.3\,$arcsec in diameter, each Starbug has a fibre exclusion radius of $10\,$arcmin, limiting the positioning of adjacent fibres. Since our survey strategy involves over 20 passes of each sky region, this limitation does not affect our scientific goals. In Section~\ref{observing}, we describe how our tiling algorithm takes this into account to produce optimal fibre configurations.

With a resolving power of $R\gtrsim2000$, TAIPAN will be capable of a wide variety of galaxy and stellar science, including distance-scale measurements to 1\%, velocity dispersions down to at least $70\,$km~s$^{-1}$, and fundamental parameters (e.g., temperature, metallicity, and surface gravity) for every bright star in the southern hemisphere. In addition to the Taipan survey described here, the TAIPAN positioner will also be used in bright time to carry out the FunnelWeb survey\footnote{\url{https://funnel-web.wikispaces.com}}, targeting all $\sim3$ million southern stars to a magnitude limit of $I_\mathrm{Vega} \lesssim 12$ over the three years from 2017-2019. The TAIPAN positioner itself also serves as a prototype for the Many Instrument Fibre System (MANIFEST) facility, which is being designed for the Giant Magellan Telescope and would operate from the mid-2020s \citep{Saunders2010,Lawrence2014}. This technology will also be used in a new multiplexed integral field spectrograph for the Anglo-Australian Telescope (AAT), Hector \citep{Lawrence2014b,Bryant2016}, which will undertake the largest-ever resolved spectroscopic survey of nearby galaxies \citep{BlandHawthorn2015}.

\section{Scientific goals}
\label{goals}

\subsection{A precise measurement of the local distance scale}
\label{goals_bao}

The present-day expansion rate of the Universe (the Hubble constant, \ho) is one of the fundamental cosmological parameters. Measuring \ho\ accurately and independently of model assumptions is a crucial task in cosmology.

\begin{figure}
 \includegraphics[trim={1.75cm 0.25cm 0.5cm 0.5cm},width=\columnwidth]{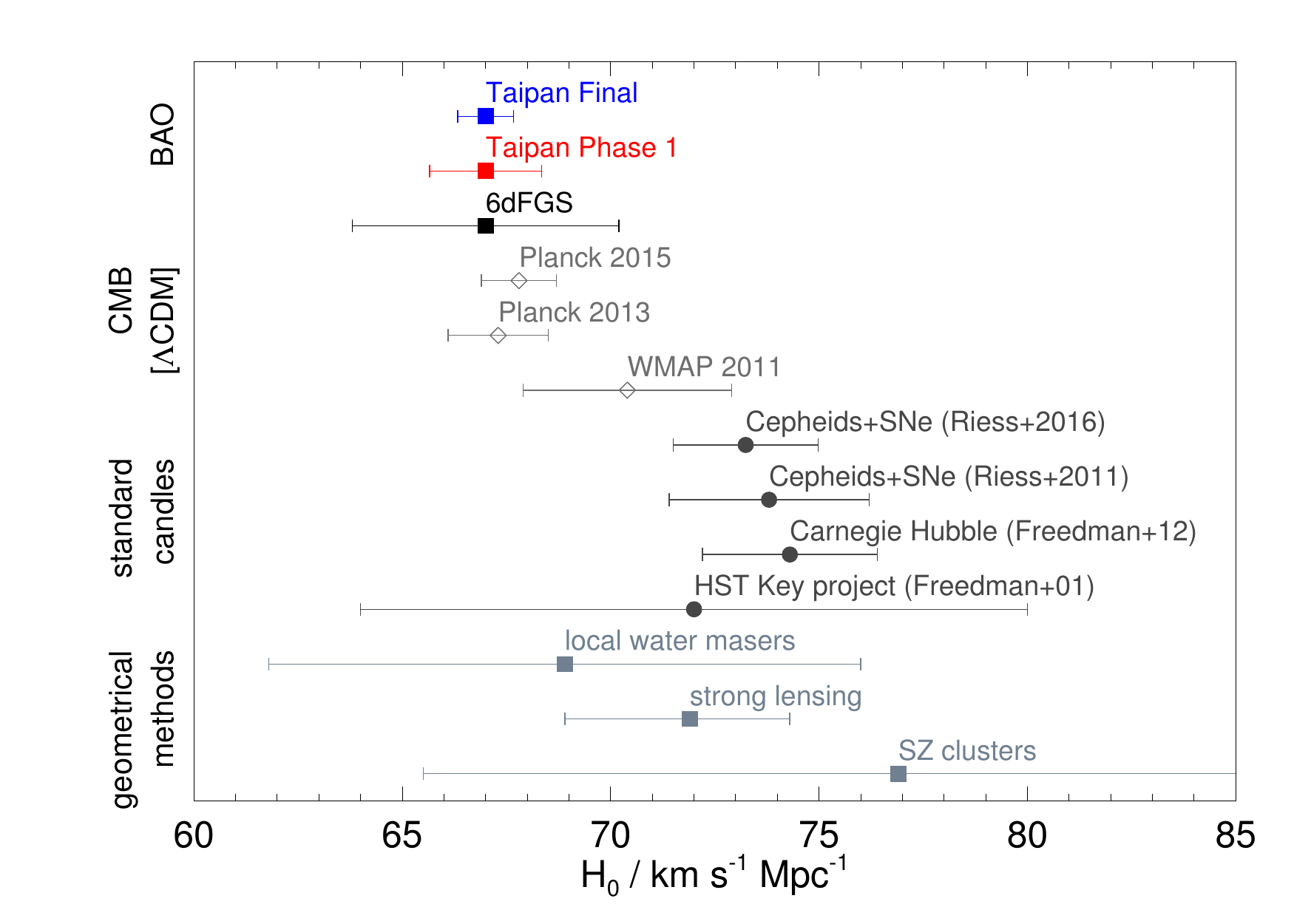}
 \caption{Measurements of the local value of the Hubble constant, \ho, from different methods and datasets.
The predictions from CMB measurements are from WMAP \protect\citep{Larson2011,Komatsu2011} and Planck \protect\citep{Planck2014}, and both are obtained assuming the $\Lambda$CDM model. Other measurements are based on local standard candles (Cepheid stars and supernovae) by \protect\cite{Riess2011,Riess2016} and \protect\cite{Freedman2001,Freedman2012}, and geometrical methods: local water masers \protect\citep{Reid2013}, strong lensing \protect\citep{Bonvin2017}, and galaxy clusters \protect\citep{Bonamente2006}. We also show the BAO peak measurement from 6dFGS \protect\citep{Beutler2011}. The forecast precision of the Taipan survey result (at the \protect\citealt{Beutler2011} 6dFGS BAO value) is also indicated, both after the $\sim1.5$ years of observations (Taipan Phase~1, in red) and after the full survey (Taipan Final, in blue).  The precision achieved with Taipan will address the current tension between measurements based on the CMB and those using standard candles.}
 \label{holocal}
\end{figure}

Current cosmological surveys, combined with high-precision measurements of the CMB \citep{Planck2015}, Type-Ia supernovae \citep{Freedman2012, Betoule2014, Riess2016}, and weak gravitational lensing \citep{Heymans2012,Abbott2016,Hildebrandt2017}, point to a consensus $\Lambda$CDM cosmological model: a spatially-flat Universe dominated by cold dark matter and dark energy, the latter having caused the late-time Universe to undergo a period of accelerated expansion. Under the $\Lambda$CDM paradigm, dark energy exists in the form of a cosmological constant, although understanding its underlying physics poses theoretical challenges (e.g.\,\citealt{Joyce2016}). One of the main goals for cosmology since the discovery of this accelerated expansion in the late 1990s \citep{Riess1998,Perlmutter1999} is explaining the nature of dark energy, and whether it is indeed a cosmological constant or a more exotic extension to the cosmological model. 
This requires precise constraints on the dark energy density, $\Omega_\Lambda$, and on the dark energy equation of state, $\omega$. A challenge in doing so is that the dark energy parameters are partially degenerate with the Hubble constant and so demand a direct and model-independent measurement of \ho. Direct measurements of the distance-redshift relation using standard candles (e.g., Cepheids and supernovae)  rely on calibrations of the distance ladder that have their own uncertainties and may suffer from systematics (see e.g.\,\citealt{Freedman2010} for a review). Importantly, there is currently significant tension between the value of \ho\ from CMB and BAO measurements at high redshift (which must assume a $\Lambda$CDM model) and low-redshift standard candle studies (e.g.\,\citealt{Riess2011,Bennett2014,Efstathiou2014,Spergel2015,Riess2016}; see Fig.~\ref{holocal}).

Taipan is designed to obtain a direct, 1\%-precision measurement of the low-redshift distance scale in units of the sound horizon at the drag epoch. Measuring the distance scale, which is governed mainly by \ho, at that precision will allow us to investigate whether the current discrepancy between low-redshift standard candle measurements and higher-redshift CMB and BAO measurements is due to systematic errors, or points to deviations from the current $\Lambda$CDM model.
We will use the imprint of baryonic acoustic oscillations (BAOs) in the large-scale distribution of galaxies as a `standard ruler' \citep{EisensteinHu1998,Colless1999,Blake2003,Seo2003,Eisenstein2005,Bassett2009}.
Pressure waves in the photon-baryon plasma prior to the epoch of recombination left an imprint in the baryonic matter after the Universe had cooled sufficiently for the photons and baryons to decouple. In response to the hierarchical collapse of dark matter, the baryons went on to form galaxies, and the remnants of these pressure waves, the BAOs, can be detected in the clustering of these galaxies. The BAO signal has a small amplitude, however, and its robust detection requires galaxy redshift surveys mapping large cosmic volumes (of order 1 Gpc$^3$) and large numbers of galaxies (over $10^5$; e.g.\,\citealt{Blake2003,Blake2006,Seo2007}). 
The sound-horizon scale has been calibrated to a fraction of a percent by CMB measurements \citep{Planck2015} and, as the BAO method utilises clustering information on large ($\sim100\,h^{-1}\,$Mpc) scales, it is robust against systematic errors associated with non-linear modelling and galaxy bias on smaller scales \citep{Mehta2011,Vargas2016}. 
Moreover, BAOs have been found to be extremely robust to astrophysical processes that can substantially affect other distance measures \citep{Eisenstein2007, Mehta2011}. 

The direct and precise low-redshift measurement that we aim to obtain with Taipan is crucial for several reasons. First, dark energy dominates the energy density of the local Universe in the standard cosmological model, and thus new gravitational physics should be more easily detectable here, than at high redshift.  Second, cosmological distances are governed by \ho\ at low redshift, implying that the usual Alcock-Paczynski effect \citep{Alcock1979} causes negligible extra uncertainty. Third, distance constraints at low redshift provide valuable extra information in cosmological fits, helping to break degeneracies between \ho\ and dark energy physics that affect the interpretation of higher-redshift distances \citep{Weinberg2013}; in particular, model predictions normalised to the CMB diverge at low redshift. Fourth, the high galaxy number density that can be mapped at low redshift, and the availability of peculiar velocities, allow for the application of multiple-tracer cross-correlations.

Thanks to their robustness, low-redshift BAO measurements provide a promising route to understanding the current tension between local measurements of \ho\ and the value inferred by the CMB, and identify whether this is due to systematic measurement errors or unknown physics. Measurements of \ho\ with of order 1\% errors from both the `distance ladder' reconstructed by standard candles, and from the `inverse distance ladder' using the CMB and BAOs will allow for strong conclusions about the nature of this disagreement (e.g., \citealt{Bennett2014}). For example, if the current disagreement remains after such precise measurements have been made, the statistical significance of this difference will then greater than $5\,\sigma$, substantially strengthening the argument for physics beyond the standard cosmological model.

\begin{figure}
 \includegraphics[trim={0cm 0.25cm 0cm 0cm},width=\columnwidth]{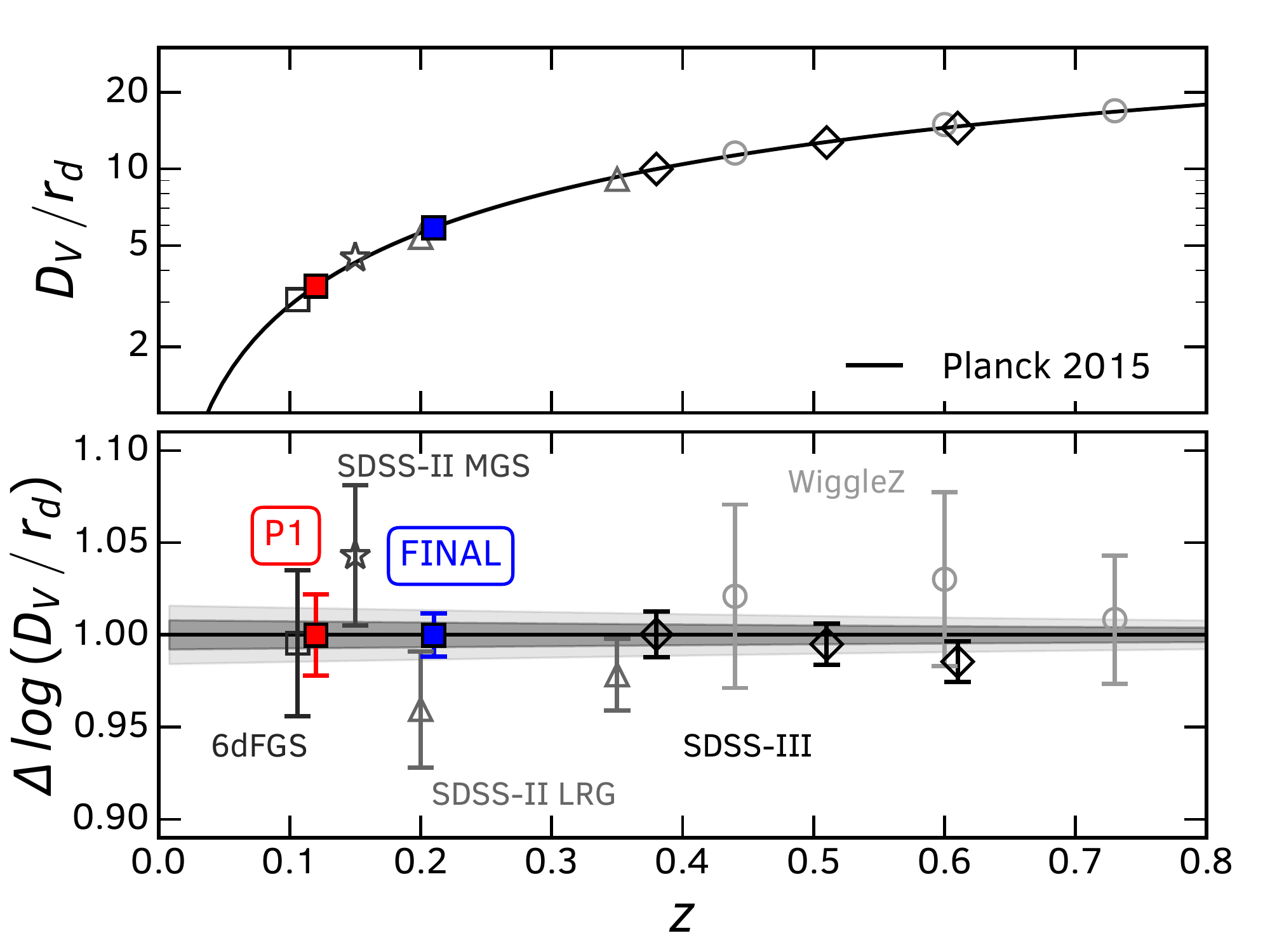}
 \caption{BAO distance-redshift measurements, expressed as $D_\mathrm{V}/r_\mathrm{d}$, the ratio of the volume-averaged comoving distance and the size of the sound horizon at recombination. Coloured filled squares show the predictions for the Taipan Phase~1 (P1, in red) and Taipan Final (in blue). The other symbols show existing measurements from the 6dFGS (open square; \citealt{Beutler2011}), SDSS-III BOSS-DR12 (diamond; \citealt{Alam2016}), SDSS-II MGS (star; \citealt{Ross2015}), SDSS-II LRG (triangle; \citealt{Percival2010,Xu2012}) and WiggleZ datasets (circle; \citealt{Kazin2014}). The lower panel shows the measured/predicted BAO scale divided by the BAO scale under the fiducial Planck cosmology, such that points in perfect agreement with \protect\cite{Planck2015} would lie on the black line. The black line and surrounding grey regions show the best-fit, $1\sigma$ and $2\sigma$ confidence regions for a $\Lambda$CDM cosmological model based on the results of \protect\cite{Planck2015}.}
 \label{bao_scale}
\end{figure}

We forecast the precision of BAO distance-scale measurements with the Taipan survey using the Fisher matrix method of \cite{Seo2007}. We assume a survey area of $2\pi$ steradians, the galaxy redshift distributions for Taipan Phase~1 and Taipan Final selections presented in Section~\ref{strategy}, a linear galaxy bias factor $b=1.2$ and the redshift incompleteness predicted by our exposure time calculator. We assume that `reconstruction' of the baryon acoustic peak \citep{Eisenstein2007} can be performed such that the dispersion in the bulk-flow displacements can be reduced by $50\%$, and combine the angular and radial BAO measurements into a single distortion parameter, which is equivalent to the measurement of a volume-weighted distance $D_V(z)$ at the survey effective redshift $z_{\rm eff}$ in units of the sound horizon $r_d$, $D_V(z_{\rm eff})/r_d$. We find that Taipan is forecast to produce a measurement, in Phase~1 and Final stages, of $D_V/r_d$ with precision $2.1\%$ and  $0.9\%$, respectively, at effective redshift $z_{\rm eff}=0.12$ and $0.21$ (covering an effective volume $V_{\rm eff} = 0.13$ and $0.59\,h^{-3}$~Gpc$^3$).
The BAO method has been widely used in the past decade to obtain robust distance measurements. Such measurements are shown in Fig.~\ref{bao_scale} for a number of large galaxy surveys, alongside predictions for Taipan.
The forecast Taipan distance-scale measurements are competitive with the best-existing constraints from other surveys.

\subsection{Detailed maps of the density and velocity field in the local Universe}
\label{peculiar}

\subsubsection{Density field and predicted peculiar velocity field} 

6dFGS mapped local, large-scale structures in
the southern hemisphere using a sample of over $125,000$ redshifts.
Taipan, with a fainter magnitude limit and improved completeness, will allow us to map the local
cosmography at greatly enhanced resolution. Observed redshift-space maps
can be transformed, via reconstruction techniques, into real-space maps
which allow the local density field to be determined (see
e.g.\,\citealt{Branchini1999,Erdogdu2006,Carrick2015}). Along with
densities, these techniques simultaneously predict the peculiar velocities of galaxies (i.e.\,
the deviations in their motions from a uniform Hubble flow).
The improved fidelity provided by the Taipan redshift
survey will yield a map of the local density field from which a detailed
prediction can be made for the local peculiar velocity field. This will
allow us to quantify the contributions from known dominant large nearby
structures (e.g.\,Great Attractor/Norma; \citealt{LyndenBell1988,Mutabazi2014}), and reach out far enough
to fully map the gravitational influence of the richest nearby
superclusters such as Shapley ($z=0.05$; \citealt{Proust2006}), Horologium-Reticulum
($z=0.06$; \citealt{Lucey1983,Fleenor2005}), and the recently-discovered Vela supercluster ($z=0.06$; \citealt{KraanKortweg2017}).

\subsubsection{Fundamental Plane peculiar velocities}
\label{pv}

Independently of density field reconstructions, the peculiar velocities of galaxies can
be determined directly from measurements of redshift-independent
distances via
\begin{equation}
v_\mathrm{pec} \approx cz - \ho\ D,
\end{equation}
where $cz$ is the redshift in km\,s$^{-1}$, \ho\ is the
local Hubble constant in km\,s$^{-1}$\,Mpc$^{-1}$, and $D$ is the
distance in Mpc (see e.g.\,\citealt{Davis2014} for the rigorous
formulation).

Four distance indicators have been used extensively in peculiar velocity
studies: the Fundamental Plane (FP), the Tully-Fisher (TF) relation,
Type Ia supernovae, and surface brightness fluctuations. Each method has advantages and limitations, in terms of sample size, intrinsic precision, and sensitivity to systematic
uncertainties. The FP and TF relations have the key advantage that they
can be applied efficiently to large numbers of galaxies.
Previous large-scale velocity surveys include the 6dFGS peculiar velocity survey (6dFGSv; \citealt{Springob2014}) using the Fundamental Plane,  and the SFI++ and 2MTF surveys \citep{Springob2007,Hong2014} using the Tully-Fisher relation. Some of these measurements are included in the compiled catalogues Cosmicflows-3 \citep{Tully2016} and COMPOSITE \citep{Feldman2010}, which incorporate measurements from several distance indicators.

The FP is the scaling relation that links the velocity dispersion,
effective radius, and effective surface brightness of early-type galaxies
\protect\citep{Dressler1987, Djorgovski1987}:
\begin{equation}
\label{eq:fp}
\log R_e = a \log{\sigma_0} + b\,\langle \mu_e \rangle + c,
\end{equation}
where $R_e$ is the effective (half-light) radius,
$\langle \mu_e \rangle$ is the mean surface brightness within $R_e$, and
$\sigma_0$ is the central velocity dispersion; $a$ and $b$ are the plane
coefficients, and $c$ is the plane zero-point. After small and
well-defined corrections, $\langle \mu_e \rangle$ and $\sigma_0$ are
effectively distance-independent quantities, whereas $R_e$ scales with
distance. Measuring the former quantities thus provides an estimate of
physical effective radius, and comparison with the measured angular
effective radius yields the angular diameter distance of the galaxy.

6dFGSv was the first attempt to
gather a large set of homogeneous FP-based peculiar velocities over the
whole southern hemisphere. It exploited velocity dispersion
measurements from the 6dFGS spectra for a local
($z$$<$0.055) sample of early-type galaxies, combined with
2MASS-based measurements of the photometric parameters, to derive
$\sim$9,000 FP distances with an average uncertainty of 26\%
\protect\citep{Magoulas2012,Campbell2014,Springob2014}.

Taipan will provide measurements for at least five times as many
galaxies as 6dFGSv, sampling the volume within $z<0.05$ more densely,
and reaching out to $z\sim0.1$. A key aspect of the
Taipan peculiar velocity work will be linking the improved predicted
peculiar velocity field derived from the redshift survey with the large
set of homogeneous FP peculiar velocities over the {\it same} local
volume. 

Taipan will also bring several substantial improvements
expected to reduce FP distance errors to $\sim20\%$. The most important
of these are:
\begin{enumerate}[topsep=0pt]
\item achieving smaller random and systematic velocity dispersion errors
 by increasing the spectral signal-to-noise
  and by taking advantage of the higher instrumental resolution of the
  TAIPAN spectrograph to measure velocity dispersions to
  $70\,$km\,s$^{-1}$ (compared to $112\,$km\,s$^{-1}$ for 6dFGSv). Taipan
  will allow us to better determine the random and systematic errors in the velocity
  dispersion by using a large number of independent repeat measurements;
  in addition there will be over 4,000 galaxies in the sample that have
  SDSS velocity dispersion measurements. This overlap sample will
  provide a robust bridge between the Taipan and SDSS datasets and allow
  us to assemble an (almost) all-sky FP-based peculiar velocity
  sample;
\item selecting early-type galaxies more efficiently by taking advantage of the higher quality (smaller PSF) and
  deeper imaging data available for the southern hemisphere from e.g.
  SkyMapper, Pan-STARRS, and Vista Hemisphere Survey (VHS; \citealt{McMahon2013});
\item improving the homogeneity of FP photometric parameters
  by combining measurements from the optical $r$$i$ bands from
  SkyMapper and Pan-STARRS, and the near-infrared bands from 2MASS and
  VHS;
\item improving the FP method precision by correcting for the contributions of stellar population
  properties (such as age and metallicity) to the intrinsic FP
  scatter (e.g.\,\citealt{Springob2012}), and by calibrating the FP from
  spatially-resolved spectroscopy (e.g.\,\citealt{Cortese2014,Scott2015}).
\end{enumerate}
Controlling and minimising the distance errors is critical to the Taipan
peculiar velocity survey strategy. The principal data requirement is sufficiently high
signal-to-noise in the optical spectra to derive a precise and robust
measurement of the stellar velocity dispersion for each galaxy, since uncertainty in
velocity dispersion measurements is the dominant source of
observational uncertainty in the distance estimates from the FP. The aim
is to make this {\em observational} uncertainty substantially
(i.e.\,at least two times) smaller than the $\gtrsim20\%$ {\em intrinsic}
uncertainty in the FP distance estimates. We therefore set the goal of
achieving a precision of $\lesssim10\%$ for the Taipan velocity
dispersion measurements. Based on previous experience in measuring velocity
dispersions in other large spectroscopic survey programmes, including
6dFGSv and SDSS, this requires obtaining a median continuum
S/N$\gtrsim15\,$\AA$^{-1}$ over the key rest-frame wavelength range from
H$\beta$ (4861\,\AA) to Fe5335 (5335\,\AA).

In total we expect Taipan to provide new high-quality FP distances for
about $50,000$ early-type galaxies with $z< 0.1$ (see Section~\ref{phase1_pv}).
Using these measurements we will robustly characterise the local velocity
field and, in combination with the redshift survey, place tighter
constraints on cosmological models.
The constraints from our Taipan FP survey will be further tightened with the addition
of TF peculiar velocities obtained by the WALLABY survey \citep{Koda2014,Howlett2017}.

\subsubsection{Testing the cosmological model with peculiar velocities}
\label{bulkflow}

The Taipan survey will enable both a definitive cosmography of the local density and velocity fields as well as precision constraints on the cosmological model. From the former, Taipan will determine in detail the structures contributing to the motion of the Local Group and the scale on which this converges to its motion with respect to the CMB. In the case of the latter, the peculiar velocities complement the redshift survey, and test the gravitational physics linking peculiar velocities to the underlying mass fluctuations, which can be modelled using linear theory and/or traced by the redshift survey.

The observed motion of the Local Group with respect to the local CMB
rest frame arises from the attraction of the entire surrounding dark matter mass
distribution. At present the main contributions are still not
well established. The scale at which these contributions
converge to the CMB dipole and amplitude of the external
bulk flow due to mass fluctuations outside the local volume remain 
matters of debate \protect\citep{Feldman2010,Lavaux2010, Bilicki2011, Nusser2011,Hoffmann2015, 
Carrick2015}. A key goal of the Taipan peculiar velocity survey is to
investigate and definitively characterise the local bulk flow.

The 6dFGS peculiar velocity survey, with $\sim9,000$ peculiar velocities, 
is the largest single survey so far undertaken to understand the origin of this
observed motion \protect\citep{Springob2014,Scrimgeour2016}. While this 
found that the statistical measurement of galaxy bulk motions in
the local Universe is consistent with predictions from linear theory (assuming the standard
$\Lambda$CDM model), there was evidence for an external bulk flow in the general
direction of the Shapley supercluster; i.e.\,a component of the bulk
flow that is not predicted by the model velocity field interior to this
volume as derived from redshift surveys
\citep{Springob2014}. By mapping the velocity field of
galaxies with better precision over a larger volume than previous
surveys (extending well beyond the Shapley supercluster and out to $z \sim 0.1$), Taipan will
measure this external bulk flow with greater precision and determine
whether it is due to the Shapley supercluster being more massive than
currently estimated, to other large structures at greater distance (e.g.\,the newly-discovered Vela supercluster; \citealt{KraanKortweg2017}), or to unexpected deviations from standard $\Lambda$CDM cosmology (e.g.\,\citealt{Mould2017}).

\begin{figure}
 \includegraphics[trim={0.5cm 0.6cm 0.25cm 0.25cm},width=\columnwidth]{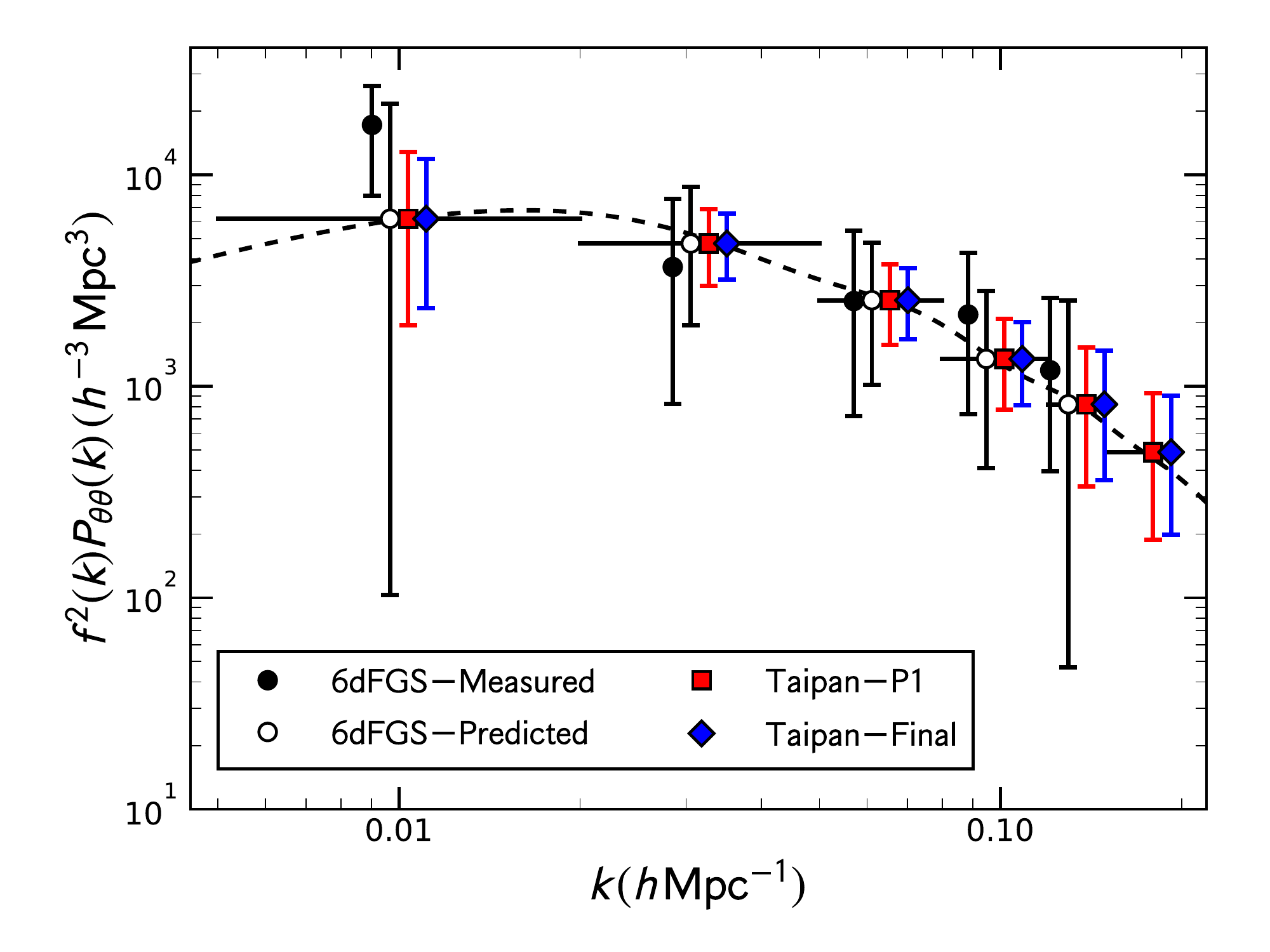}
 \caption{Measurements and predictions for the {\em scale-dependent} growth rate (in distinct $k$-bins) multiplied by the velocity divergence power spectrum for our fiducial cosmology, using {\em only} the peculiar velocity samples of 6dFGS and Taipan. For 6dFGS, we plot both the measurements from \cite{Johnson2014} and forecasts as solid and open points. The dashed line shows the prediction from GR. The predictions/measurements are sensitive to the power averaged across each bin (solid horizontal lines), but the placement of the points within each bin is arbitrary. There is some discrepancy between the 6dFGS measurements and forecasts, but in all bins we see significant improvement in Taipan over the 6dFGS predictions, which we expect to translate through to the measurements made with Taipan. Hence Taipan will allow us to place tight constraints on the scale-dependence of the low redshift growth rate, which is an important test of GR.}
 \label{vel_power}
\end{figure}

The volume and sample size provided by the Taipan peculiar velocity survey will also allow, in principle, the measurement of the bulk flow as a function of scale not just in a single volume around the Local Group, but in tens of independent volumes on scales up to $\sim$100\,Mpc/$h$. A more effective way to capture this information is through the galaxy velocity power spectrum. This was computed directly by \cite{Johnson2014} using 6dFGSv (see also \citealt{Macaulay2012} for a similar parametric analysis). With a larger volume and denser sampling of the velocity field, the Taipan peculiar velocity survey will provide a much more precise velocity power spectrum over a wider range of scales, as shown in Figure~\ref{vel_power}. This improved velocity power spectrum will yield improved constraints on specific cosmological parameters that are degenerate when only the galaxy density power spectrum is available (see \citealt{Burkey2004,Koda2014}).
In terms of constraining the cosmological model, the key advantages of peculiar velocities are that: (i)~they trace the gravitational physics on very large scales that are not accessible by standard redshift-space distortions from galaxy redshift surveys, where modified gravity scenarios often show interesting deviations; (ii)~the correlated sample variance between the peculiar velocities and density fields allows some quantities to be constrained with errors below the sample-variance limit; and (iii)~the availability of both velocity and density field data is critical for marginalising over relevant nuisance parameters that would otherwise impair redshift-space distortion fits. These issues are explored in relation to the Taipan survey by \cite{Koda2014}  and \cite{Howlett2017}. 

\subsection{Testing models of gravity with precise measurements of the growth rate of structure}
\label{gravity}

One possible explanation for the apparent `dark sector' of the Universe, and for the tensions between our current cosmological model and observations, is a modification to Einstein's theory of General Relativity \citep{Einstein1916}. A key observable that can be used to distinguish between models of gravity is the growth rate of structure, defined as $f=d \ln g / d \ln a$, where $g$ is the linear perturbation growth factor, and $a$ is the expansion factor. This growth rate defines how fast galaxies fall into gravitational potential wells, and governs the peculiar velocities that we measure. The growth rate as a function of redshift can be parameterized as $f(z) = \Omega_m(z)^{\gamma}$, where $\Omega_m$ is the matter density of the Universe, and $\gamma$ depends on the physical description of gravity  (e.g.\,\citealt{Wang1998,Linder2005,Weinberg2013}). General Relativity in a $\Lambda$CDM model predicts $\gamma = 0.55$ \citep{Linder2007}. Therefore, by measuring the growth rate and, in particular, constraining $\gamma$, we can test models of gravity.

Taipan will measure the growth rate of structure in two complementary ways. First, the statistical correlations between the measured peculiar velocities and the density field traced by the redshift survey can be used to constrain the growth rate with a particular sensitivity to large-scale ($>100 h^{-1}$~Mpc) modes (as described above; Fig.~\ref{vel_power}). 
Such measurements were made previously using the COMPOSITE and 6dFGSv samples \citep{Macaulay2012, Johnson2014}, but our survey will improve on these by providing over five times more peculiar velocities.

The second probe of the growth of structure is using the redshift-space clustering of galaxies. The peculiar motions of galaxies change the amplitude of clustering in an anisotropic way, an effect called `redshift-space distortions' (RSD; \citealt{Kaiser1987}). Galaxies infalling towards structures along the line-of-sight will appear further away or nearer than they truly are when their distance is inferred from their redshift. On the other hand, infall perpendicular to the line-of-sight will not change the measured redshift from the value based on its true distance. Hence, otherwise isotropic distributions of galaxies appear anisotropic, and the clustering amplitude of the galaxies changes depending on the angle we look at compared to the line-of-sight. Additionally, averaging over all lines-of-sight no longer gives the same clustering as if the galaxies had zero peculiar velocity.

RSDs are a powerful probe of the growth rate of structure and have been used in many large galaxy surveys. However, galaxies are biased tracers of the underlying density field that influences their peculiar motions, and in the redshift-space clustering of galaxies, there is a strong degeneracy between the effects of galaxy bias and RSD. Measurements of the clustering of galaxies from their redshifts alone is also limited by cosmic variance. One of the greatest advantages of the Taipan survey comes from combining the large number of redshifts that can be used to measure the effects of RSD \textit{and} the direct measurements of the peculiar velocities. The combination of these has the ability to break the degeneracy with galaxy bias and overcome the limits of cosmic variance \citep{Park2000,Burkey2004,Koda2014,Howlett2017}. Moreover, direct peculiar velocities and RSD are sensitive to large and intermediate scales, respectively, allowing any scale-dependent modifications to the growth rate to be mapped out (Fig.~\ref{vel_power}).

\begin{figure}
\centering
\includegraphics[trim={0.5cm 1cm 0.2cm 0.2cm},width=\columnwidth]{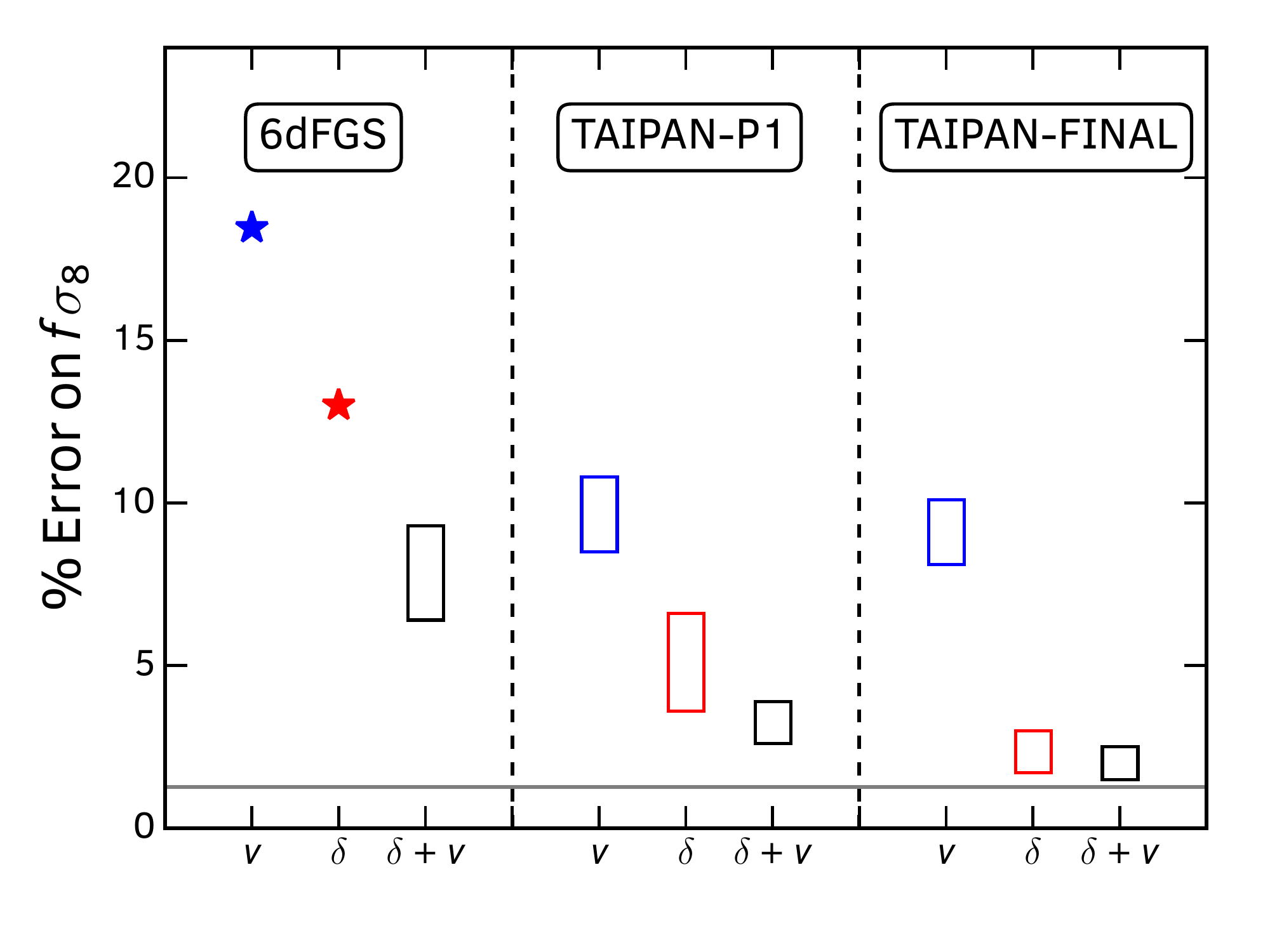}
  \caption{Measurements/predictions of the percentage error in the growth rate\protect\footnotemark using the peculiar velocity (denoted $v$) and redshift (denoted $\delta$) samples of the 6dFGS and Taipan surveys separately and in combination. This highlights how a relatively small number of peculiar velocity measurements can be used to improve upon the constraints on the growth rate from redshift surveys alone. Stars are measurements for the 6dFGS from \protect\cite{Beutler2012} and \protect\cite{Johnson2014}. All other points are predicted error regions, produced using the method of \cite{Howlett2017} assuming different levels of prior knowledge on any nuisance parameters. The upper edge of each region assumes no prior knowledge, while the lower edge assumes perfect knowledge. The horizontal line corresponds to the uncertainty from \protect\cite{Planck2015} at $z=0$ assuming $\Lambda$CDM and General Relativity. With the Taipan survey we constrain the growth rate almost as tightly as Planck, but, crucially, without requiring the assumption of $\Lambda$CDM.}
  \label{growth2}
\end{figure}

The effect of combining RSD and direct peculiar velocities is demonstrated in Figure~\ref{growth2}, where we compare the percentage error on measurements of the growth rate we expect to obtain with Taipan alongside existing and predicted constraints from the 6dFGS \citep{Beutler2012,Johnson2014}. These forecasts were produced using the method detailed in \cite{Koda2014} and \cite{Howlett2017}. In all cases we see a marked improvement on the growth rate constraints when the two probes of the growth rate are combined, compared to their individual constraints. In particular, we expect Taipan Phase~1 and Taipan Final to constrain the growth rate to $4.5\%$ and $2.7$\% precision, respectively.

\footnotetext{When measuring the growth rate using RSD in the two-point clustering, the galaxy bias, growth rate $f$ and $\sigma_8$ (the root-mean-square of mass fluctuations on scales of $8\,h^{-1}\,$Mpc) are completely degenerate on linear scales. We chose $f\sigma_8$ as our forecast parameter because this combination allows us to test different gravity theories without prior knowledge of the galaxy bias and $\sigma_8$, as shown in \protect\cite{Song2009}.}

\begin{figure}
\centering
\includegraphics[trim={0.5cm 1cm 0.2cm 0.2cm},width=\columnwidth]{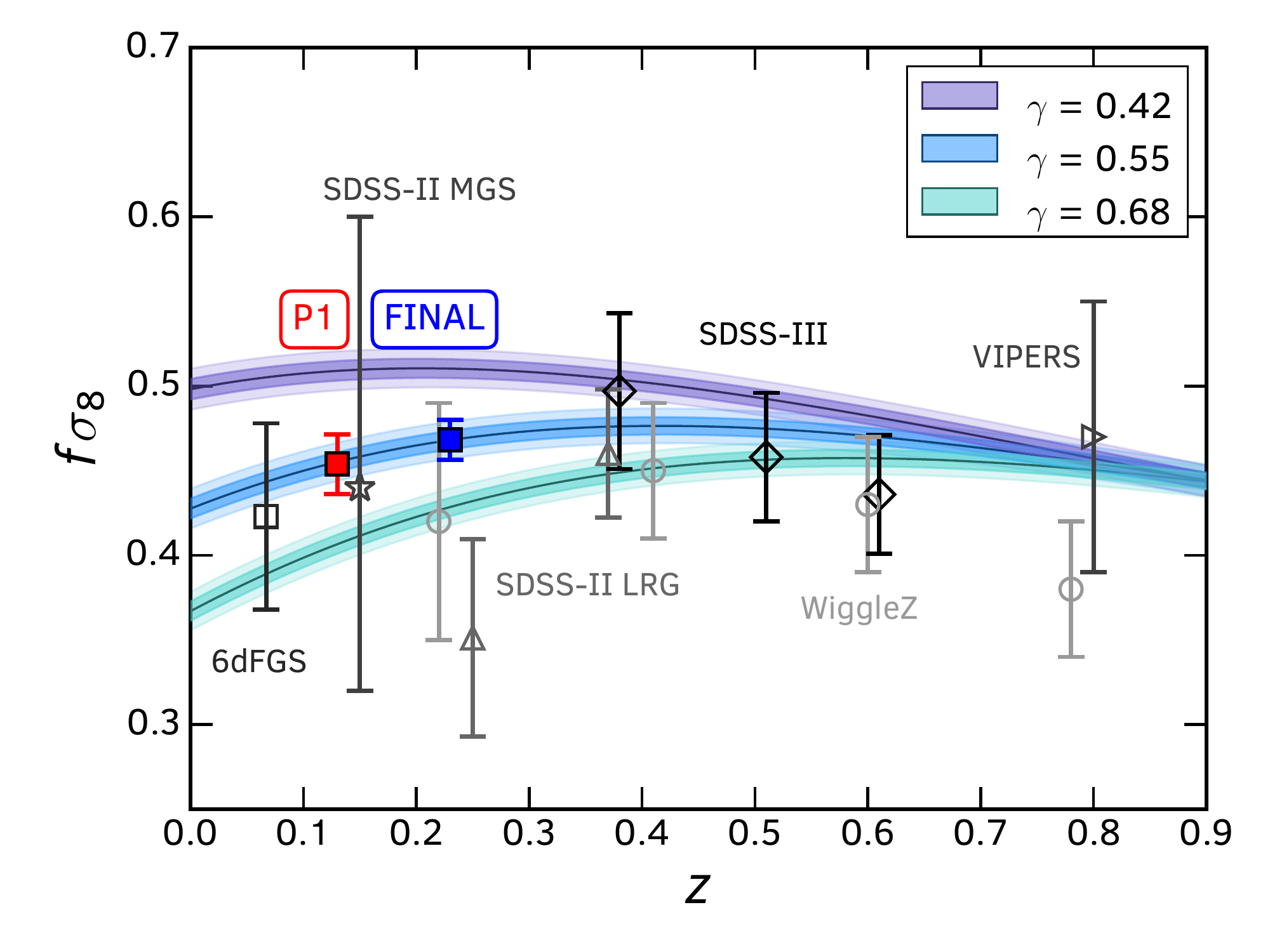}
  \caption{A comparison of different measurements and predictions of the growth rate of structure, $f\sigma_8$, as a function of redshift, from various galaxy surveys. Coloured points (filled squares) show the predictions for Taipan Phase~1 (in red) and Final (in blue). Other points are existing measurements from the 6dFGS (open square; \citealt{Beutler2012}), SDSS-III BOSS-DR12 (diamond; \citealt{Alam2016}), SDSS-II MGS (star; \citealt{Howlett2015a}), SDSS-II LRG (triangle; \citealt{Samushia2012}) and WiggleZ datasets (circle; \citealt{Blake2011}). The coloured bands indicate the growth rate obtained for different theories of gravity using the parameterisation of \protect\cite{Linder2007} and assuming a flat-$\Lambda$CDM cosmology based on the results of \protect\cite{Planck2015}. The value $\gamma=0.55$ corresponds closely to the prediction from General Relativity for $\Lambda$CDM. This demonstrates that a precise measurement at low redshift such as the one enabled with Taipan will distinguish between different models of gravity.}
  \label{growth}
\end{figure}

To highlight the strong constraining power of Taipan, we show the predictions for the growth rate alongside measurements using RSD from other large galaxy surveys in Figure~\ref{growth}. Taipan measurements utilising both RSD and peculiar velocities are expected to significantly improve over measurements from current surveys and are well-placed in a regime where we expect large relative deviation between different gravity models.

Furthermore, modified gravity models rely on \textit{screening mechanisms} that allow deviation from general relativity in under-dense regions, making cosmic voids particularly useful to probe gravity (e.g.\,\citealt{Achitouv2016}). With Taipan, the complete and dense mapping of local large-structure will allow us to define an exquisite sample of voids, and the surrounding redshift-space distortion will provide the best measurement of the linear growth rate in under-dense regions. 
The local Universe is particularly relevant for testing non-standard dark energy theories that dominate the late-time cosmic expansion.  Current constraints on the linear growth rate around voids have been performed at low redshift with the 6dFGS dataset in \cite{Achitouv2017} and at higher redshifts with SDSS \citep{Hamaus2016} and the VIMOS Public Extragalactic Redshift Survey (VIPERS; \citealt{Hawken2016}).
The Taipan sample can also be used to test gravitational physics by performing cross-correlations with overlapping weak lensing and CMB datasets.
  
\subsection{The lifecycle of baryons as a function of mass and environment}
\label{galaxies}

\begin{figure}
\centering
\includegraphics[trim={0.1cm 0.2cm 0.1cm 0cm},width=\columnwidth]{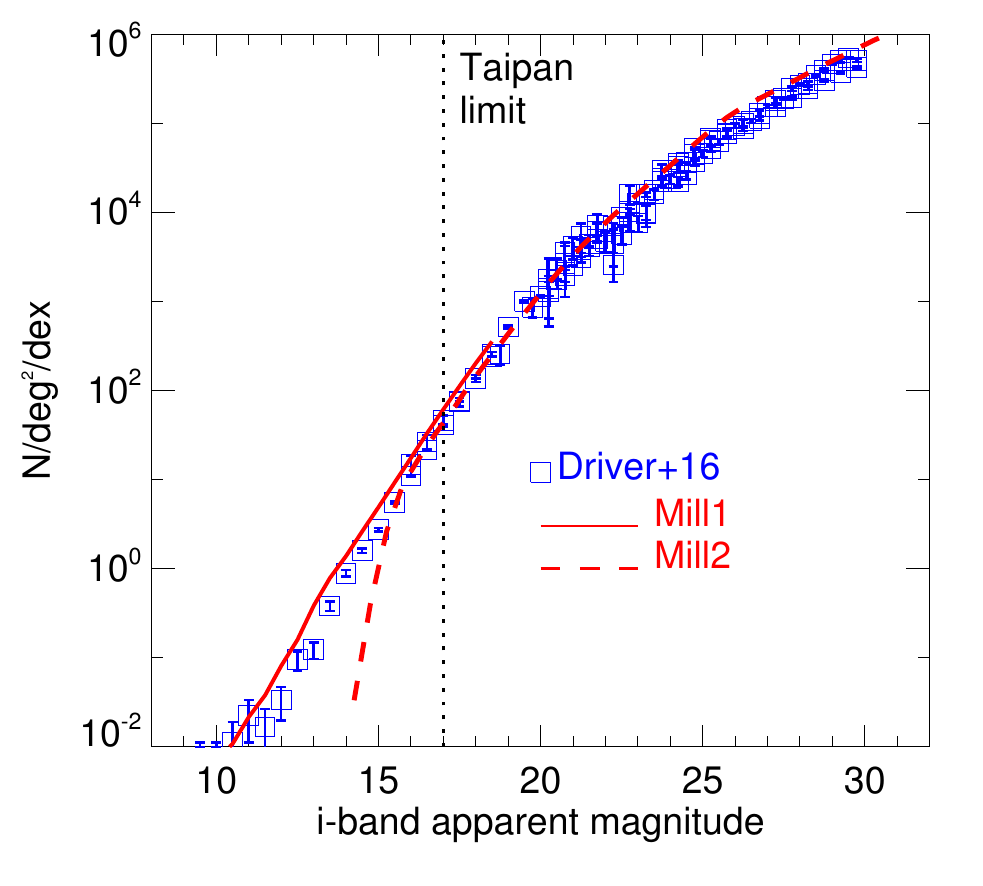}
  \caption{Comparison between the observed local $i$-band number counts from the recent compilation of \protect\cite{Driver2016} (blue squares), and the predictions from the {\sc galform} semi-analytic galaxy formation model (\citealt{Bower2006,Lagos2012}; red lines). The solid and dashed lines correspond to lightcones generated from the Millennium 1 \citep{Springel2005} and Millennium 2 \citep{Boylan2009} cosmological runs, respectively. The vertical dotted line shows the Taipan magnitude limit, $i=17$. The two different Millennium realisations combined provide precise results over a large range of scales (enabled by the significantly better spatial and mass resolution of the Millennium 2 run compare to Millennium 1, which includes a larger volume); this ensures that galaxies are resolved in the full stellar mass range from $10^6$ to $10^{12}$~\msun.}
  \label{iband_counts}
\end{figure}

Previous spectroscopic galaxy surveys at low redshifts, in particular SDSS ($z\simeq0.1$; \citealt{Abazajian2009}) and GAMA ($z\simeq0.2$; \citealt{Driver2011,Liske2015}), have provided a wealth of information on the properties of present-day galaxies and the physical processes affecting their evolution. However, many questions remain regarding the dominant processes responsible for quenching star formation in galaxies (e.g.\,\citealt{Baldry2004,Blanton2009,Schawinski2014}). These open questions include: what are the roles of interactions, the large-scale environment, and active galactic nuclei (AGN) in quenching star formation? What drives the efficiency of star formation? And how do the properties of the neutral gas reservoir in galaxies relate to the star-forming properties? A way to address these issues is through a comprehensive sample of local galaxies spanning a wide range of environments, with large enough sample sizes to isolate the effects of different physical processes and characterise rare populations, such as galaxies rapidly transitioning from star-forming to quiescent. Wide multi-wavelength coverage is also needed to optimally trace all the baryons in galaxies, including stellar populations of different ages, neutral and ionised gas in the interstellar medium (ISM), and dust.
Taipan will address crucial questions in galaxy evolution by capitalising on a few key advantages over existing spectroscopic surveys at low redshift.

Taipan has two main advantages over SDSS. First, since Taipan is a multi-pass survey, there will be many opportunities to revisit targets affected by `fibre collisions' i.e., the inability to simultaneously observe targets that are too close on the sky plane.
This will allow us to identify close pairs of galaxies (with separations smaller than the $55\,$arcsec limit imposed by fibre collisions in a given SDSS plate; \citealt{Strauss2002,Blanton2003}), to study the effect of close interactions and mergers, and measure the environment density and halo masses (e.g.\,\citealt{Robotham2011,Robotham2014}). Second, Taipan will overlap with the WALLABY HI survey\footnote{\url{http://www.atnf.csiro.au/research/WALLABY/}} \citep{Koribalski2012}, carried out with ASKAP \citep{Johnston2008}, which aims to cover three-quarters of the sky and expects to detect $\sim500,000$ galaxies in HI (e.g.\,\citealt{Duffy2012}). Thanks to this overlap, we will characterise the neutral gas reservoir of an unmatched number of optically-detected galaxies spanning a wide range of halo masses, stellar masses, and environments. At the same time, Taipan will provide the stellar and halo mass measurements to contextualise the HI data from WALLABY. Taipan will also be competitive with the deeper and spectroscopically-complete GAMA survey in the low-redshift regime, thanks to the much larger sky coverage (about $20,600$~deg$^2$ for Taipan versus $286$~deg$^2$ for GAMA), which implies a volume sampled by Taipan at $z<0.1$ of $1.5\times10^8$~Mpc$^3$, i.e.\,72 times larger than the volume sampled by GAMA in the same redshift range ($2.13\times10^6$~Mpc$^3$)\footnote{\url{http://cosmocalc.icrar.org}; \cite{Robotham2016}.}. We note that GAMA cosmic variance is estimated to be $\sim13\%$ \citep{Driver2010}, which will be reduced to about $5\%$ for the final Taipan survey.

To predict the properties of our magnitude-limited sample, we use a mock catalogue of galaxies extracted from a state-of-the-art theoretical galaxy formation model. We extract $2,600$~deg$^2$ lightcones from the \cite{Lagos2012} version of the {\sc galform} semi-analytic model \citep{Cole2000,Bower2006}, which includes the post-processing of the Millennium $N$-body $\Lambda$CDM cosmological simulation \citep{Springel2005,Boylan2009}. Figure~\ref{iband_counts} shows that the model successfully reproduces the observed $i$-band counts from \cite{Driver2016} over a wide range of magnitudes. The version of {\sc galform} implemented by \cite{Lagos2012} is ideal for our purposes because it not only reproduces the observed optical properties of local galaxies, but also gas properties such as the local HI and H$_2$ mass functions \citep{Lagos2011b}, thanks to a sophisticated treatment of the two-phase (i.e.\,atomic and molecular) neutral ISM based on an empirical, pressure-based star formation law \citep{Blitz2006}.\footnote{The Taipan and WALLABY lightcones presented here are available upon request (via claudia.lagos@icrar.org).}

\subsubsection{Galaxy pairs and the close environments of galaxies}

\begin{figure}
\centering
\includegraphics[trim={0.2cm 0.5cm 0.2cm 0cm},width=0.45\textwidth]{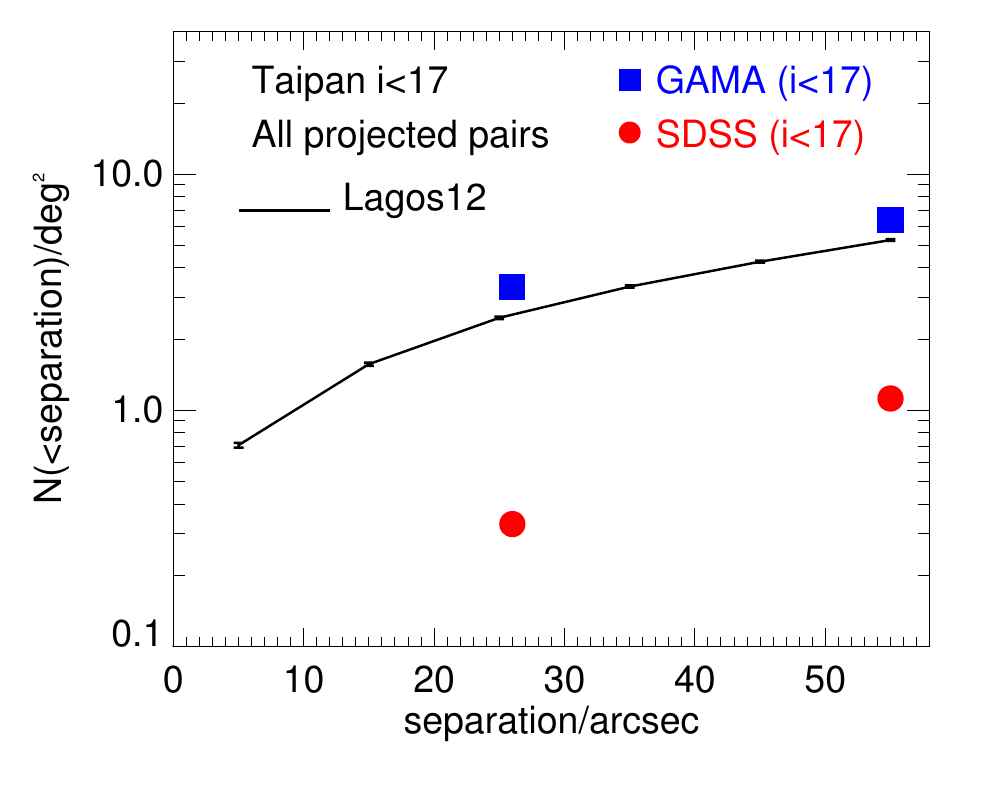}
 \caption{Cumulative number density of galaxy pairs as a function of sky separation. The black lines show the predictions for Taipan ($i<17$) based on the {\sc galform} model \citep{Lagos2011a,Lagos2012}. Poisson errors are of the order of the thickness of the line. The blue and red squares show the number density of pairs detected by GAMA and SDSS, respectively, at 25 and 55~arcsec separations (using the same magnitude limit).}
 \label{taipan_pairs}
\end{figure}

Most galaxies do not evolve in isolation. Galaxy interactions and mergers are theoretically predicted to have an important role in the $\Lambda$CDM hierarchical view of galaxy evolution (e.g.\,\citealt{Barnes1992,Hopkins2010}). Observationally, both the small-scale and large-scale environments of galaxies have been shown to have an impact on their properties, such as their morphology, star formation and AGN activity, and stellar mass growth (e.g.\,\citealt{Dressler1980,Postman1984,Kauffmann2004,Alonso2006,Bamford2009,Ellison2008,Ellison2010,Scudder2012,Wijesinghe2012,Brough2013,Robotham2014,Aspaslan2015,Gordon2017} and references therein). Despite the large advances in this field enabled by modern spectroscopic and imaging surveys, it is challenging to disentangle the effects of close interactions from the large-scale environment, and the intrinsic properties of the galaxies, e.g., stellar masses, gas content, and existence of an AGN (e.g.\,\citealt{Blanton2005,Ellison2011,Scudder2015}).

To quantify merger/interaction rates, and their large-scale environment, we must be able to identify close pairs of galaxies, i.e., we need a highly complete spectroscopic survey (e.g.\,\citealt{Robotham2011,Robotham2014}). The main limitation of SDSS in this field is the inability to account for galaxy pairs with a projected sky separation smaller than $55\,$arcsec due to fibre collisions \citep{Strauss2002}. This biases galaxy pairs identified with SDSS towards large separations, with less than 35 per cent of photometrically-identified galaxy pairs in the SDSS spectroscopic sample having separations less than $55\,$arcsec \citep{Patton2008}. Taipan will mitigate this problem by visiting each field in the sky multiple times to achieve very high ($>98\%$) spectroscopic completeness down to $i=17$.

In Figure~\ref{taipan_pairs}, we use the \cite{Lagos2012} model to predict the number of close pairs expected with Taipan. Taipan Final will detect about $140,000$ galaxy pairs at separations closer than 55~arcsec (i.e.\,$\sim54$~kpc at $z\simeq0.05$), and about $70,000$ pairs with sky separations less than 25~arcsec (i.e.\,$\sim27$~kpc at $z\simeq0.05$). This is about 10 times more pairs than those detected by SDSS over a similar area and magnitude limit. Taipan will detect a similar surface density of pairs as GAMA (at the same magnitude limit), but with the advantage of sampling a much larger volume.
The significantly larger statistical sample produced by Taipan will allow us to dissect the pair sample into various properties. We will measure pair fractions in the local Universe as a function of stellar mass ratio, primary (and satellite) mass and morphology, and larger-scale environment, expanding the previous GAMA study by \cite{Robotham2014}, thus obtaining a rich low-redshift baseline for studies of the evolution of interactions and mergers with cosmic time (e.g.\,\citealt{XuZhao2012}) that is less affected by cosmic variance than GAMA.

By combining our pair dataset with multi-wavelength surveys (e.g., in the radio), we will investigate how close pairs affect the properties of galaxies, such as their star formation and AGN activities.
For example, features seen at kpc-scales in the radio jets of AGN may be generated or influenced by galaxy pairs. In particular, it has been posited that radio galaxies showing distorted and twisted lobes, the so-called `bent-tail galaxies', arise in the presence of close pairs in which the gravitational interaction of the pair provides a mechanism to twist the radio jet \citep{Begelman1984}, although this is just one of several mechanisms that may be responsible for radio jet morphology. While evidence of the optical host of a bent-tail galaxy being part of a pair has been found in small samples of nearby objects (e.g., \citealt{Rose1982,Mao2009,Pratley2013,Dehghan2014}), there has been no systematic large-scale study of the topic to date. The combination of Taipan with recent and anticipated southern radio surveys will provide the first opportunity to address this question.

In addition to the analysis of the role of galaxy pairs, we will use a number of other environmental metrics for exploring the significance of
environment in moderating galaxy evolution. Metrics we anticipate using in the analysis of the Taipan data include the commonly used nth-nearest
neighbour approaches (e.g., \citealt{Gomez2003,Brough2013}), cluster-centric distances (e.g., \citealt{Owers2013}), galaxy groups defined using friends-of-friends
algorithms (e.g., \citealt{Robotham2011}) and lower density `tendril' structures \citep{Alpaslan2014}. Each of these metrics has advantages and disadvantages. Generally, the simpler techniques (such as nth nearest neighbour) are easier to measure for a larger fraction of a sample, but are less directly sensitive to the true underlying local environmental density. Using this broad range of metrics we can compare Taipan results directly with other published work using common measurements, and can also begin to link the metrics being used with the true physical environments in order to explore their impact on galaxies.

 Crucially, the overlap with the WALLABY survey (Section~\ref{wallaby}) will allow us to compare the atomic (HI) gas content of galaxies in pairs with that of isolated galaxies while controlling for the large-scale environment. For example, we will be able to test if pairs have an HI excess due to being associated with (invisible) gas streams from the cosmic web, or if, on the contrary, they are more HI-deficient because of interaction shocks and/or harassment dynamics, and whether these properties change as a function of environment density.

\subsubsection{Complementarity with WALLABY}
\label{wallaby}

\begin{figure*}
\centering
\begin{minipage}{0.75\linewidth}
\includegraphics[width=0.5\textwidth]{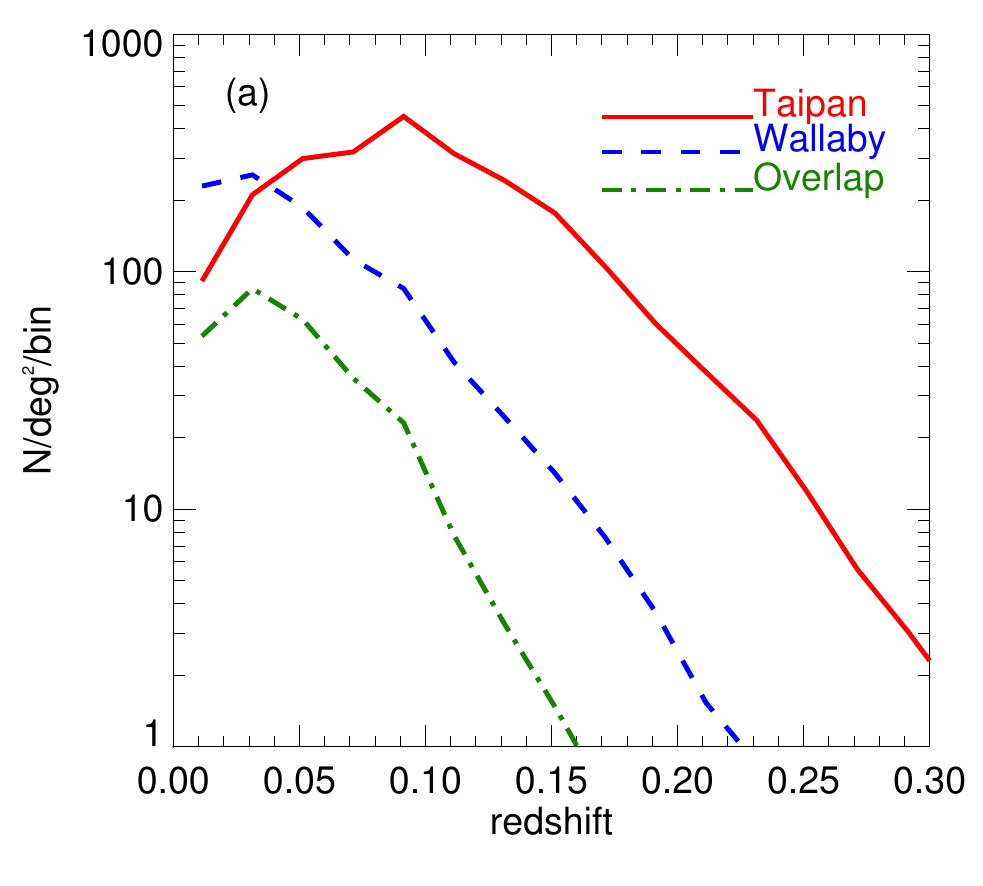}
\includegraphics[width=0.5\textwidth]{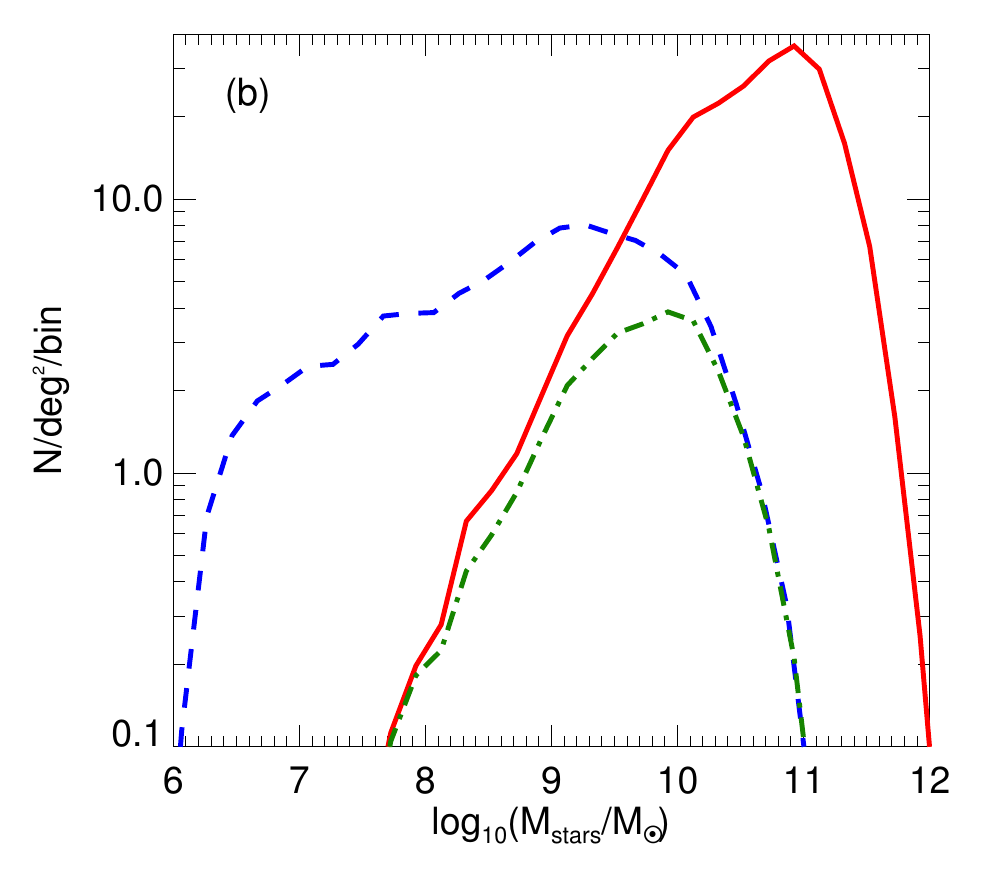}
\includegraphics[width=0.5\textwidth]{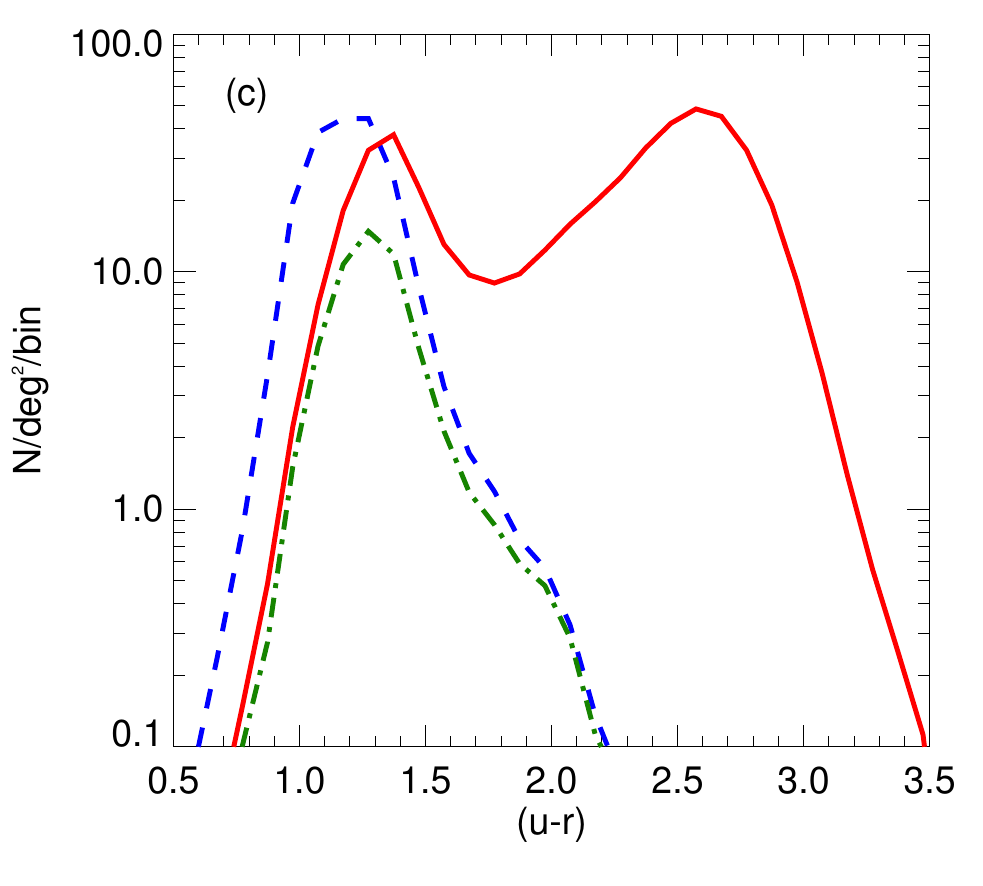}
\includegraphics[width=0.5\textwidth]{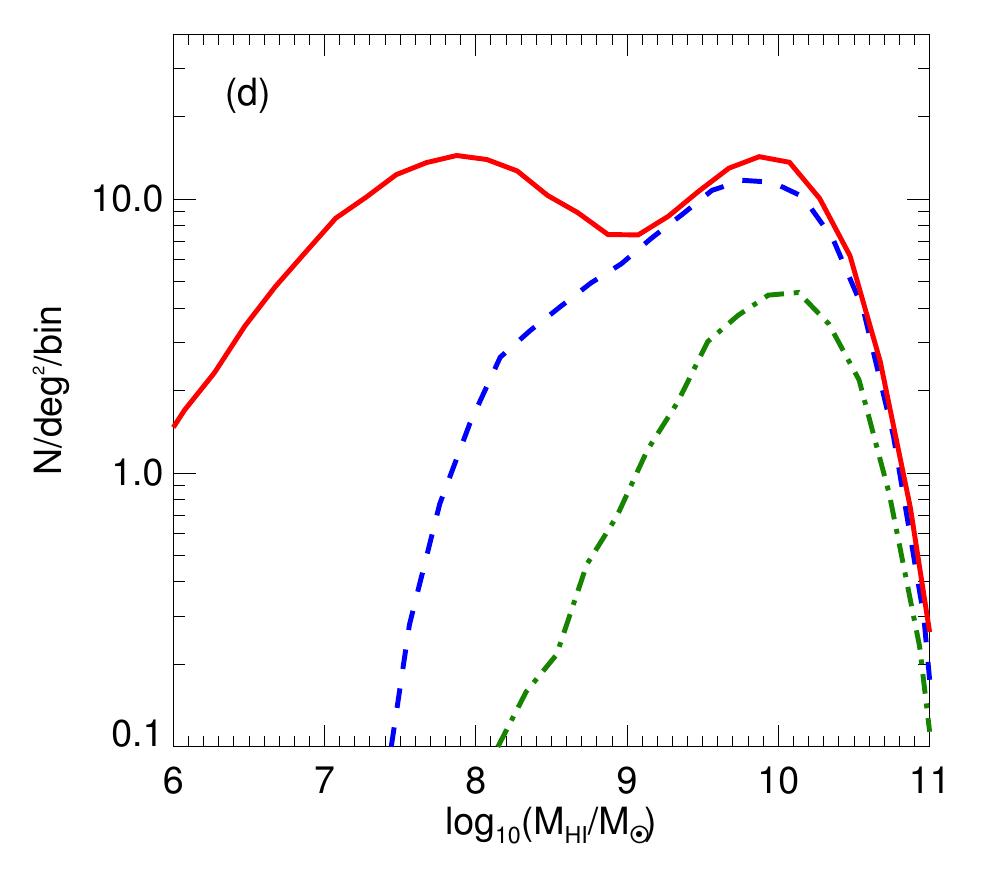}
\end{minipage}
 \caption{Predicted distribution of the properties of galaxies detected in Taipan (i.e.\,galaxies with $i\le17$; in red) and in WALLABY (i.e~galaxies with HI line detections above $5\sigma\simeq8\,$mJy, with $\sigma=1.592\,$mJy per $3.86\,$km/s velocity channel; in blue, dashed; e.g.\,\citealt{Duffy2012}) from the {\sc galform} model \citep{Lagos2011a,Lagos2011b,Lagos2012}: {\bf (a)} redshift; {\bf (b)} stellar mass; {\bf (c)} $u-r$ (rest-frame) colour; {\bf (d)} neutral hydrogen mass. The green dot-dashed lines show the distribution of properties of galaxies detected in both surveys.} 
 \label{taipan_wallaby}
\end{figure*}

The gas content of galaxies (i.e.\,the fuel for star formation) plays a crucial role in their evolution. The WALLABY survey on ASKAP will measure the HI masses for the largest ever sample of galaxies in the local Universe. Combined with Taipan (and ancillary multi-wavelength surveys), these observations will allow us to trace the evolution of the full baryonic content of galaxies as a function of mass and environment.

We use our mock galaxy catalogue to predict the properties of galaxies observed with Taipan and WALLABY based on the observational constraints of those surveys. We take `Taipan detections' to be all galaxies with $i\le17$, and `WALLABY detections' to be all galaxies with $z<0.26$ and HI line detections above 8~mJy. According to these simulations, WALLABY will obtain about $600,000$ 5-$\sigma$ HI detections over its total sky coverage of $30,940$~deg$^2$, of which $\sim140,000$ will also be Taipan targets (i.e.\,in the `overlap sample'). In Figure~\ref{taipan_wallaby}, we show the properties (redshifts, stellar masses, optical $u-r$ colours, and HI masses) of Taipan detections, WALLABY detections, and the overlap sample of Taipan+WALLABY detections. The galaxies in the overlap sample will be typically star-forming, at $z\simeq0.05$, with blue colours, stellar masses typically between $10^9$ and $10^{11}$\,M$_\odot$, and high ($>10^9$~M$_\odot$) HI masses.
With Taipan we will also be able to push down the HI mass function by stacking faint WALLABY detections.
As shown in Fig.~\ref{taipan_wallaby}(c), we expect to individually detect HI in galaxies in the green valley and blue cloud with WALLABY, but we will miss a large number of red sequence galaxies. The Taipan optical redshift information will be used to perform spectral HI stacking (e.g.\, \citealt{Delhaize2013}) for galaxies split by different properties, and as a function of distance to galaxy cluster centre (or galaxy group centre), optical colour, stellar mass, and more.

Using the Taipan+WALLABY sample we will map out how the population density from the star-forming cloud to the red sequence depends on environment, stellar mass, and gas mass. This will provide a diagnostic of the timescale of gas loss and star formation rate decline as a function of environment and mass, which can be compared to theoretical models describing ram-pressure stripping, thermal evaporation and tidal starvation in groups and clusters of galaxies (e.g.\,\citealt{Boselli2006}) .

A further application will be a stringent test of the cosmic web detachment model (\citealt{Aragon2016}; see also \citealt{Kleiner2017}). If galaxies are attached to the cosmic web and accreting gas from filaments \citep{Keres2005}, this will be reflected in their observable properties, such as their HI content. In particular, when non-linear interactions sever the link between a galaxy and the cosmic web, we will be able to directly detect the quenching taking place within these galaxies \citep{Kleiner2017} by comparing the HI-to-stellar mass ratio (i.e.\,the HI fraction) as a function of their large-scale environment.

\subsubsection{Complementarity with other multi-wavelength surveys}

A major advantage of Taipan will be the overlap with other surveys of the southern sky across various wavelengths. Taipan will provide spectroscopic redshifts for low-redshift sources detected in various continuum surveys, and we will use multi-wavelength information provided by ancillary surveys to obtain a more complete physical understanding of Taipan galaxies. In this section, we provide a non-exhaustive overview of those ancillary surveys.

In the radio, Taipan will overlap with the Evolutionary Map of the Universe (EMU) survey \citep{Norris2011} carried out on ASKAP, which will obtain the deepest, highest-resolution radio continuum (at $1.1-1.4\,$GHz) map of the southern sky.
While EMU will detect AGN to very high redshift, the bulk of its detected sources will be star-forming galaxies at fairly low redshift.
EMU is expected to detect Milky Way-type disk galaxies out to $z\sim0.3$, and simulations suggest that millions of galaxies will be detected to $z\leq0.5$. The Taipan+EMU sample of nearby galaxies, therefore, will be both larger and deeper than the Taipan+WALLABY sample. Cluster science is an important focus of EMU, particularly the detection of extended emission from galaxy clusters without selection effects. Taipan's ability to provide redshifts, and hence cluster detection and characterisation in the nearby universe, will complement this aspect of EMU.
At lower frequencies, the Galactic Extragalactic All-sky Murchison Widefield Array (GLEAM) survey \citep{Wayth2015} will provide additional AGN and interstellar medium diagnostics, as well as a complementary probe of environment through galaxy clusters \citep{Bowman2013}. In particular, despite its low resolution, the very high low-surface-brightness sensitivity of the MWA \citep{Hindson2016} combined its low-frequency capability, makes it ideal to detect older, diffuse radio plasma from AGN that are no longer active (e.g.\,\citealt{HurleyWalker2015}), as well as vastly increasing the detection of rare examples of disk-hosting galaxies with large-scale double radio lobes (Johnston-Hollitt et al.~subm., Duchesne et al.~in prep.). Thus, GLEAM will provide diagnostics for over $300,000$ active AGN and, when combined with Taipan, will also provide the rare opportunity to identify and study the optical properties of galaxies in which the AGN has been extinguished, and to examine instances in which spiral and lenticular galaxies host low-power, large-scale, double-lobed AGN. 

Taipan will be highly complementary to photometric surveys in the near-infrared to ultraviolet probing the emission by stellar populations and ionised gas in galaxies, as well as attenuation from dust in the interstellar medium, and allowing new cosmological tests.   
Surveys such as the Dark Energy Survey (DES; \citealt{DES2016}) and the VST Kilo-Degree Survey (KiDS; \citealt{deJong2017}) will obtain weak lensing maps of the southern sky. Overlapping lensing and redshift surveys are complementary because they allow new types of scientific analyses, and also mitigate systematic errors afflicting both probes \citep{Joudaki2017}. Combining Taipan with these surveys will enable joint analyses to test gravitational physics, such as the `gravitational slip' \citep{Daniel2008}, and to do cross-correlation analyses between lensing galaxies measured by Taipan and background DES/KiDS galaxies. As for systematics, lensing data allows direct measurements of galaxy bias (the main systematic affecting redshift-space distortions), and redshift data allows tests of intrinsic alignment models (which are a systematic affecting lensing analyses).

In the near-infrared, VHS \citep{McMahon2013} will enable morphological classification as well as reliable stellar mass estimates through probing the low-mass stars in Taipan galaxies. The Wide-field Infrared Explorer (WISE) all-sky survey \citep{Wright2010} also probes the stellar mass in its shorter wavelength filters (e.g.\,\citealt{Cluver2014}), while the mid-infrared filters sample the emission of polycyclic aromatic hydrocarbon (PAH) features and dust emission of Taipan galaxies, enabling studies of dust-obscured star formation and AGN activity. Deep and reliable optical photometry of the whole southern sky will soon become available through the SkyMapper Southern Survey (\citealt{Keller2007}; Wolf et al., in prep.) and the Pan-STARRS survey \citep{Kaiser2010,Chambers2016}. Ultraviolet emission is available through the Galaxy Evolution Explorer (GALEX) all-sky survey \citep{Martin2005}. Combining multi-wavelength information from these surveys will allow a complete characterisation of the physical properties of Taipan galaxies (star formation rate, stellar mass, dust content) through modelling of their spectral energy distributions (e.g.\,\citealt{daCunha2008,daCunha2010,Chang2015}).

Finally, in the X-rays, the eROSITA space telescope \citep{Merloni2012} to be launched soon, will provide an all-sky survey in the energy range up to 10 keV. This survey will provide another diagnostic of AGN activity in low-redshift galaxies observed with Taipan, as well as complementary measurements of large scale structure and environment through the detection of hot gas in galaxy clusters and groups in the X-rays.

With over one million galaxy redshifts, Taipan will provide a valuable legacy database of optical spectroscopy for galaxies over the whole southern sky, enhancing all of these other surveys.

\section{Survey design and implementation}
\label{strategy}

The Taipan galaxy survey and the FunnelWeb stellar survey will both be carried out in parallel using a new UKST+TAIPAN autonomous observing system, controlled by a `virtual observer' software package ({\em Jeeves}) developed in conjunction with AAO. This system will be responsible for planning each night's observing (including deciding which fields and targets to observe and when; see Section~\ref{observing}) and then executing each night's plan (including taking science and calibration frames, as well as managing the telescope by, for example, opening or closing the dome in the case of bad weather or twilight).
Observing time will be split between two surveys, with the FunnelWeb stellar survey undertaken when the Moon is above the horizon, and the Taipan galaxy survey done when the Moon is below the horizon.
Once data is acquired, it will be processed by a custom data processing pipeline, then archived and later disseminated through a public database.

In this section, we describe the implementation of the Taipan galaxy survey, from the science-driven target selection (Sections~\ref{target_selection}, \ref{phase1}, \ref{final}), to the automated observing (Section~\ref{observing}), processing (Section~\ref{data_processing}) and archiving of the data (Section~\ref{archiving}). In Section~\ref{ancillary}, we describe plans for a Taipan priority and ancillary science programme complementary to the main galaxy survey.

\subsection{Science-driven survey implementation}
\label{target_selection}

To achieve its scientific goals, Taipan will obtain optical spectra for a magnitude-limited ($i\le17$) sample of galaxies with near-total completeness across the whole southern sky. This will be supplemented by a `luminous red galaxy' (LRG) sample for high-precision BAO measurements (Section~\ref{final}) to a fainter magnitude limit ($i=18.1$).
Based on preliminary estimates of the TAIPAN throughput (Fig.~\ref{throughput}) our required S/N targets for  spectral measurements (e.g.\,redshifts, velocity dispersions, emission line fluxes) necessitate a minimum integration time of 15\,min per object, to which will be added repeat 15\,min visits to build up S/N as needed (for example, for peculiar velocity targets; Section~\ref{phase1_pv}).

Taipan requires a reliable input photometric catalogue providing the optical magnitudes of galaxies brighter than $i=18.1$ across the whole southern sky. The currently ongoing SkyMapper Southern Survey\footnote{More information and data releases can be found at \url{skymapper.anu.edu.au}.} (\citealt{Keller2007}; Wolf et al., in prep.) images the southern hemisphere in $uvgriz$ filters \citep{Bessell2011} and is the ideal and natural choice for Taipan target selection, since it will provide reliable and deep optical photometry. However, due to the unavailability of sufficiently deep SkyMapper data over the hemisphere at the start of Taipan observations in late-2017, we take a two-phase approach:
\begin{enumerate}[topsep=0pt]
\item {\em Taipan Phase~1} (from late-2017 to end of 2018). For the first phase of the survey, we have devised an observing strategy that will enable us to make a start on our three main scientific goals: measurement of large-scale structure across a large effective volume at $z \lesssim 0.2$; measurement of peculiar velocities for a large number of $z\lesssim0.1$ early-type galaxies via the FP; and demographic studies of galaxy properties as a function of halo and stellar mass and environment.
Each of these three science projects have different data requirements that are not always well-aligned -- where the first two strongly prefer a wide area, the second demands near-total completeness. Accordingly, we have identified three subsamples from already available input photometric catalogues to prioritise in the first phase of the survey, which we describe in Section~\ref{phase1}.
\item {\em Taipan Final} (from the early 2019). We expect that TAIPAN will have been upgraded to 300 Starbugs and that deep SkyMapper data will be available by the beginning of 2019. Therefore, for the second phase of the survey, we will select our targets using SkyMapper, with the goal being to obtain a near spectroscopically complete sample down to $i=17$ (i.e. similar to SDSS), along with the supplementary LRG sample needed to achieve our target 1\%-precision BAO distance measurement (Section~\ref{final}).
\end{enumerate}
We note that, while this two-phase approach is driven by the availability of input photometric catalogues, and of the upgrade to 300 fibres, our strategy allows us to maximise the early scientific return of Taipan, while ensuring the Taipan Phase~1 sample is effectively contained within the Taipan Final sample.
In Table~\ref{tab:surveys}, we summarise the main properties of the Taipan survey, and put it in context with other wide-area spectroscopic galaxy surveys.

\subsection{Taipan Phase~1}
\label{phase1}

\subsubsection{2MASS-selected sample for BAO science}

\begin{figure}
 \includegraphics[trim={0cm 0cm 0cm 0cm},width=\columnwidth]{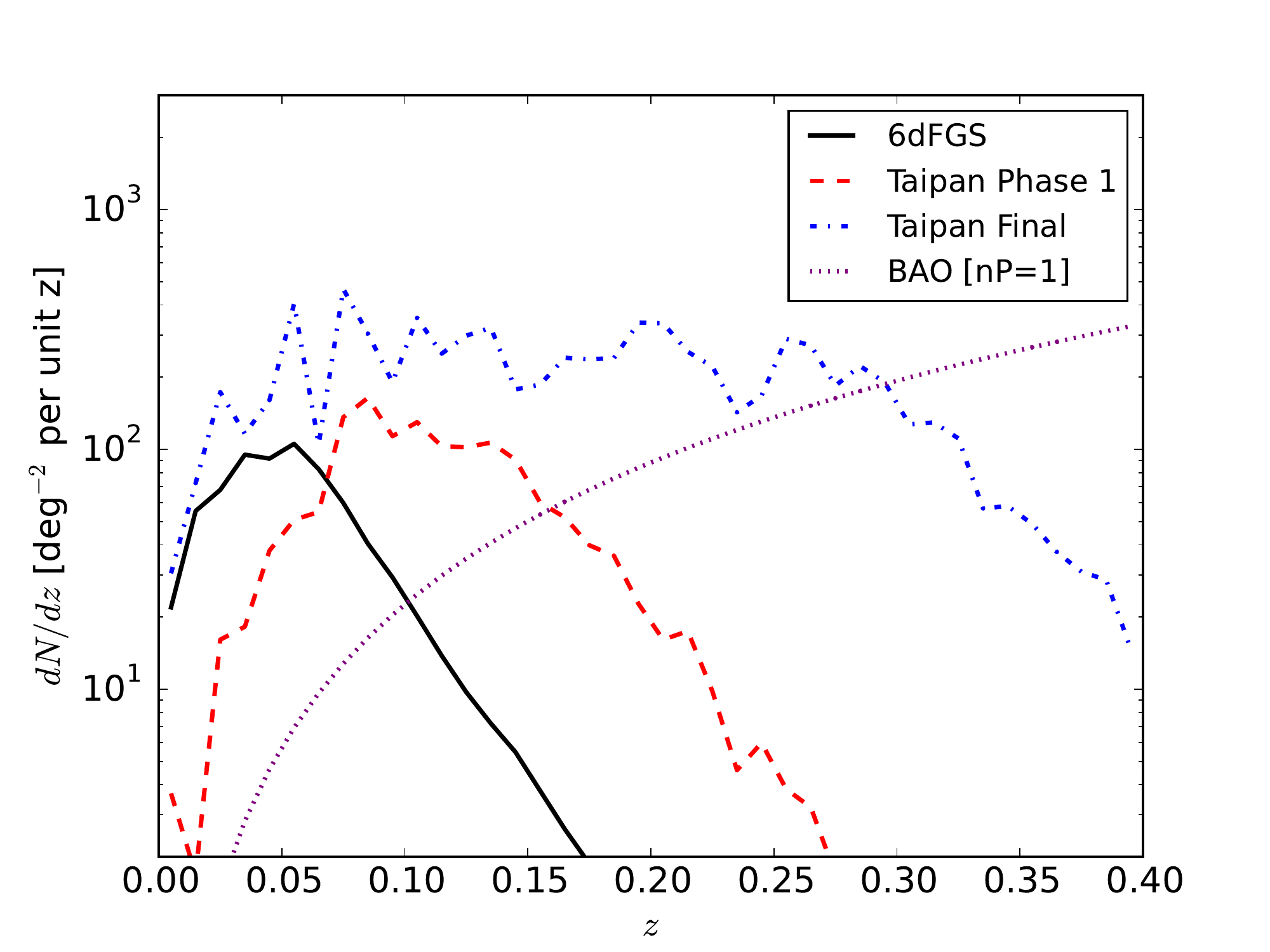}
 \caption{Predicted redshift distribution of the Taipan Phase~1 (in red) and Final (in blue) BAO sample, compared to 6dFGS (solid black line). The dotted purple line shows the number density needed to balance cosmic variance and shot noise in clustering measurements.}
 \label{redshift_nz}
\end{figure}

In terms of survey design, the requirement for the best possible measurement of the BAO distance scale is to achieve the largest possible survey volume. In the first instance, this means using the widest possible survey area, so that the precision of the measurement is principally determined by the redshift distribution of the target population.
As shown in Figure \ref{redshift_nz}, the ideal case is to have a tracer population that uniformly samples the survey volume, which naturally prefers high-redshift targets over lower-redshift ones. 
BAO science can also tolerate a low completeness.

All these factors motivate our strategy to prioritise near-infrared selected targets from the 2-Micron All Sky Survey (2MASS; \citealt{Skrutskie2006}).
The 2MASS near-infrared photometry is stable, well calibrated, and well understood across the full sky. 
We select galaxies with $J_\mathrm{Vega} < 15.4$, and with near-infrared colour $J-K > 1.2$ (Vega), which ensures that most targets will also satisfy the Taipan Final $i\le17$ selection, while being an efficient way of isolating 2MASS sources with the highest value for cosmological science in Taipan Phase~1 (Fig.~\ref{redshift_nz}).

To identify and select as many galaxy targets at the highest redshifts as possible, we supplement the 2MASS Extended Source Catalogue (XSC; \citealt{Jarrett2000}) with targets selected from the 2MASS Point Source Catalogue (PSC; \citealt{Cutri2003}).
Using the PSC photometry, we select galaxy targets on the basis of (i) their $J-K$ colour; and (ii) the difference between their 4-arcsec aperture fluxes and point-source-profile-fit fluxes, to select more extended objects. 
Through comparison with SDSS star/galaxy identifications in an overlapping field, we find that our selection criteria excludes 99.96\% of SDSS-identified stars, and retains 96.8\% of SDSS-identified galaxies.
While this efficiently selects galaxies at $0.07<z<0.15$ that are absent from the XSC, the result is only a relatively modest ($\lesssim10\,$\%) increase in our number of targets.

Using this sample, we predict that the Taipan Phase~1 2MASS-selected sample will map almost $300,000$ redshifts over an
effective volume $V_{\rm eff}=0.13\,h^{-3}\,$Gpc$^3$, obtaining a distance-scale error of $2.1\%$ at an effective redshift $z_{\rm eff}=0.12$ (Section \ref{goals_bao}).

\subsubsection{6dFGS-selected sample for peculiar velocity science}
\label{phase1_pv}

Taipan peculiar velocity science takes advantage of TAIPAN's wide-area and multi-fibre capabilities to survey a substantial number of galaxy peculiar velocities over a large volume. Precision and homogeneity are the other key considerations motivating the peculiar velocity science survey requirements, as highlighted in Section~\ref{peculiar}. The aim of observing a large number of new distance and peculiar velocity measurements in the local universe will be supported by ensuring Taipan observations have a sufficiently high signal-to-noise-ratio to derive a precise and robust velocity dispersion for these nearby early-type galaxies. The improved resolution of the TAIPAN spectrograph is one of the main improvements over the 6dFGS peculiar velocity survey. In addition, the Taipan survey will incorporate many independent repeat measurements to determine systematic errors, will access higher quality imaging data for visual classification, and will use deeper multi-band imaging for deriving homogeneous FP photometric parameters (Section~\ref{pv}).

To achieve these goals, the observing strategy for the peculiar velocity galaxies is to revisit each target until a S/N of $15\,\AA^{-1}$ is attained. Based on the expected performance of UKST+TAIPAN, we estimate that, with the selection criteria described below, this S/N threshold can be achieved for almost all Phase~1 peculiar velocity targets in four or fewer visits (i.e. $15-60$\,min total integration time). This number of visits is feasible because the Taipan BAO sample density is high enough that the survey will need to re-visit each field up to 20 times. The number of visits needed for peculiar velocity targets imposes a significant observational cost on the survey; it is therefore critical to pre-select these peculiar velocity targets as efficiently as possible. 

In Taipan Phase~1, we take advantage of the fact that 6dFGS obtained spectra for $\sim125,000$ galaxies with $K_\mathrm{Vega}<12.75$ 
and $\delta<0\deg$ (out of which $9,000$ galaxies already have velocity dispersions and FP distance measurements from 6dFGSv; \citealt{Campbell2014,Springob2014}). These spectra allow us to identify galaxies suitable for FP distance measurements before the Taipan survey starts. We have identified approximately $40,000$ targets that, based on the 6dFGS spectra, have redshifts $z < 0.1$ and no (or weak) emission lines, and are thus potentially suitable peculiar velocity targets for Taipan. We refine the selection of these targets further by performing visual inspection with the aid of available southern hemisphere imaging data, excluding galaxies with apparent spiral features, prominent dust lanes or bars, or photometry effected by stellar contamination or interacting galaxies. This visual inspection excludes around $20\%$ of the potential targets as unsuitable for FP analysis, and results in a much cleaner sample of $33,000$ peculiar velocity targets for Taipan Phase~1.

\begin{figure}
 \includegraphics[trim={0cm 0cm 0cm 0cm},width=\columnwidth]{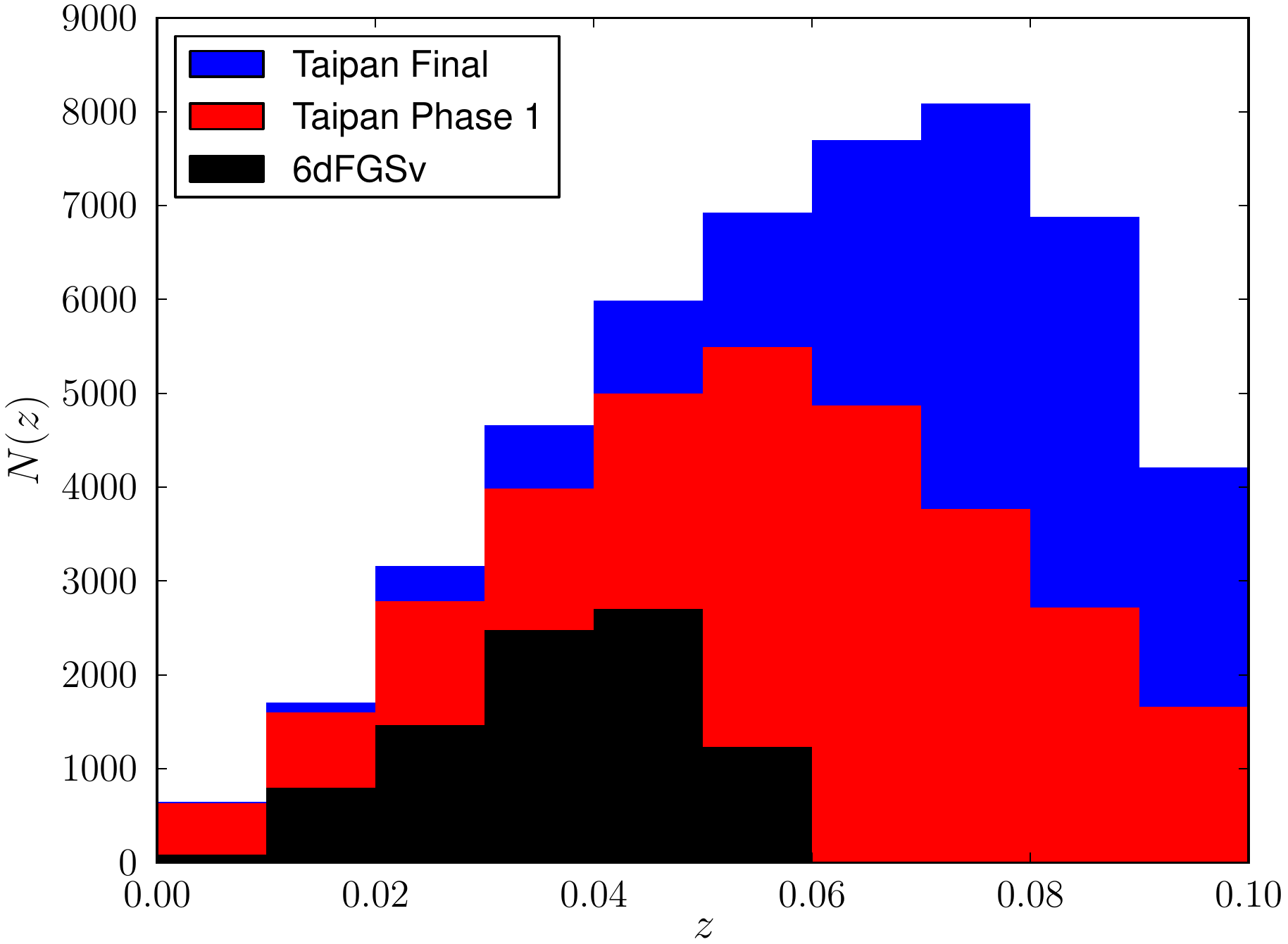}
 \caption{Comparison between the redshift distribution of the original 6dFGS peculiar velocity sample in black ($N=8,877$), Taipan Phase~1 in red ($N=33,000$) and Final in blue ($N=50,000$; predicted using selections discussed in Section~\ref{final}) velocity samples. The samples are binned in redshift intervals of $\Delta z = 0.01$.}
 \label{pv_nz}
\end{figure}

These 6dFGS-selected targets will generally be the brightest and nearest of the galaxies in the Taipan Final peculiar velocity sample. This is advantageous for the Taipan Phase~1 observations, as it means that we will be observing the easiest and highest-value (i.e. lowest peculiar velocity error) targets first. By prioritising these 6dFGS-selected targets, we expect that Taipan Phase~1 will produce FP distance measurements, and thus peculiar velocities, for a sample of up to $33,000$ local ($z < 0.1$) galaxies over the whole southern hemisphere (excluding $|b|<10^\circ$); the expected redshift distribution is shown in Fig.~\ref{pv_nz}. This already represents a factor of more than three increase over 6dFGSv, the largest existing single sample of peculiar velocities. This sample will include galaxies from 6dFGSv and other previous FP studies, providing invaluable repeat observations for probing potential systematic effects associated with differences between instrumentation and data processing.

\subsubsection{Complete sample for galaxy evolution science in the VST KiDS regions}

The third key science application for the Taipan galaxy survey is demographic studies of galaxy properties as a function of mass and environment, to derive basic empirical insights into (and quantitative constraints on) the processes that drive and regulate galaxy formation and evolution (Section~\ref{galaxies}).
While the two scientific goals described above prefer the widest possible survey area, this science application requires near total spectroscopic redshift completeness to enable robust characterisation of the immediate environments of galaxies.
To balance these competing requirements, our strategy is to prioritise $>98\%$ redshift success over a sizeable area (rather than full hemisphere) in Phase~1.

The most important factor in deciding which area, or areas, of sky to prioritise for galaxy formation and evolution science is the availability of ancillary data.
In particular, we require high quality multi-wavelength imaging and photometry, which (together with spectroscopic redshifts from Taipan) will allow us to derive stellar masses, rest-frame colours, effective radii, structural properties, etc.
A natural choice are the two VST Kilo-Degree Survey (KiDS; \citealt{deJong2017}) regions: a $780$-deg$^2$ equatorial region across the Northern Galactic cap, and a $720$-deg$^2$ Southern field around the Galactic Pole. As well as deep and very high quality $ugri$ optical imaging from VST-KiDS (which is continuing to be collected), there is already similarly good $ZYJHK$ near-infrared data from the VISTA VIKING survey \citep{Arnaboldi2007}. Furthermore, there is significant overlap with SDSS in the North, and with 2dFGRS in the South, which provide literature redshifts for $\sim90\%$ and $\sim45\%$ of Taipan targets in these two fields. Since KiDS imaging is not yet complete, we are selecting targets for Taipan Phase~1 from the slightly shallower VST-ATLAS survey,\footnote{\url{http://astro.dur.ac.uk/Cosmology/vstatlas/}} which we re-calibrate to match stellar photometry from Pan-STARRS.  

The target density of our $i\le17$ sample is $\sim60\,$deg$^{-2}$. We therefore expect a complete sample of up to $\sim90,000$ galaxies across a combined area of $1500\,$deg$^2$ in Taipan Phase~1. We also intend to prioritise other particularly interesting fields (e.g., the SPT deep field, and WALLABY early science fields) for early completeness through the course of the survey.

\begin{table*}
 \caption{Comparison between Taipan (anticipated) and other recent low-redshift, wide-area spectroscopic surveys.}
\footnotesize
   \label{tab:surveys}
 \begin{tabular}{llccccc}
  \hline
  \hline
  & & Magnitude & Number & Median redshift, & Sky area & Volume at $z_\mathrm{med}$ \\
   & & limits & of galaxies & $z_\mathrm{med}$ & /\,deg$^2$ & /\,$h^{-3}$Mpc$^3$ \\
 \hline
 \hline
  & BAO & $J_\mathrm{Vega} < 15.4$ & $3.0\times10^5$ & $0.110$ & $20,600$ & $2.0\times10^8$ \\
  &  & $J_\mathrm{Vega} - K_\mathrm{Vega} > 1.2 $ & & & & \\
  {\bf Taipan Phase~1} & Peculiar velocities & $r_\mathrm{fibre}<17.6$ & $3.3\times10^4$ & $0.055$ & $17,000$ & $2.2\times10^7$\\
  (Section~\ref{phase1}) & & & & & & \\  
  & $i$-selected & $i\le17$ & $9.0\times10^4$ & $0.086$ & $1,500$ & $7.2\times10^6$\\
   \hline
   & BAO & $i\le17$ & $2.0\times10^6$ & $0.170$ & $20,600$ & $1.3\times10^9$\\
   & &  LRG: $17<i<18.1$, $g-i>1.6$ & & & & \\
 {\bf Taipan Final} & Peculiar velocities & $r_\mathrm{fibre}<17.6$ & $5.0\times10^4$ & $0.065$ & $20,600$ & $4.3\times10^7$\\
  (Section~\ref{final}) & & $g_\mathrm{fibre}-r_\mathrm{fibre}>0.8$  & & & & \\
  & $i$-selected & $i\le17$ & $1.2\times10^6$ & $0.086$ & $20,600$ & $9.8\times10^7$\\
  \hline
  {\bf 6dFGS} & & $K_\mathrm{Vega} \le 12.65$ & $1.3\times10^5$& $0.053$& $17,000$& $2.1\times10^7$ \\
  \citep{Jones2009} &  & & & & & \\
  \hline
  {\bf 2dFGRS} & & $b_\mathrm{J} \le 19.45$ & $2.2\times10^5$& $0.110$& $1,600$& $1.7\times10^7$ \\
 \citep{Colless2001} & & & & & & \\
    \hline 
 {\bf SDSS-DR7} & & $r \le 17.77$ & $9.3\times10^5$& $0.100$& $9,380$& $7.6\times10^7$ \\
   \citep{Abazajian2009} & & & & & & \\ 
   \hline
  \hline  
\end{tabular}
\\
\flushleft{\em Note:} For Taipan Phase~1 and Taipan Final, we divide the survey into three samples: `BAO' is the redshift sample for BAOs/cosmology science, which includes the magnitude-limited sample and LRG extension; `Peculiar velocities' refers to the peculiar velocity sample; and `$i$-selected' refers to the spectroscopically-complete, magnitude-limited ($i\le17$) sample that will be used for galaxy evolution science.\\
\end{table*}

\subsection{Taipan Final}
\label{final}

We plan for survey operations to move from Phase~1 to Final at the beginning of 2019, when the SkyMapper photometric catalogues will allow us to select sources directly based on their optical magnitudes, and when the TAIPAN upgrade to 300 fibres is expected to be completed.

The final Taipan footprint will cover $2\pi$ steradians (i.e.\,$\sim 20,600$~deg$^2$), achieved through a survey area defined by $\delta\lesssim+10\deg$, $|b| > 10\deg$, and $E(B-V)<0.3$. We have chosen the survey boundaries to ensure a $2\pi$-steradian survey area, but there is some scope to expand the footprint (by $\sim10\%$) either by pushing closer to the Galactic plane, or slightly further north. 

The Taipan Final sample will comprise:
\begin{itemize}[topsep=0pt]
\item a spectroscopically-complete, magnitude-limited ($i\le17$) sample (total $\sim1.2\times10^6$ galaxies), and 
\item an LRG extension to higher redshifts needed to achieve our target 1\%-precision BAO distance measurement, with $17 < i < 18.1$ and $g-i>1.6$ (total $\sim0.8\times10^6$ galaxies).
\end{itemize}
In Fig.~\ref{redshift_nz}, we show the predicted redshift distribution for the Taipan Final sample (i.e.\,magnitude-limited sample plus LRG extension). With this sample we forecast a BAO distance measurement with $0.9\%$ precision at effective redshift $z_\mathrm{eff}=0.21$, covering effective volume $V_\mathrm{eff}=0.59\,h^{-3}\,$Gpc$^3$ (Section~\ref{goals_bao}).

The Taipan Final peculiar velocity sample will probe fainter in magnitude, while remaining within the redshift limit $z<0.1$, by identifying suitable extra targets using spectral information from the redshift survey observations of Taipan Phase~1. The Taipan Final peculiar velocity sample will adopt the same basic target selection strategy as Phase~1, selecting galaxies that are close enough to obtain reliable distance estimates ($z< 0.1$), spectra indicating little or no star-formation ($g-r>0.8$ and no strong emission lines), continuum S/N suitable for measuring a velocity dispersion in at most four visits (corresponding to $r$-band magnitudes within the fibre aperture brighter than 17.6), and velocity dispersions greater than $70$\,km\,s$^{-1}$ (to the precision this is measurable from the initial visit). We will use deep optical and infrared imaging from SkyMapper and VHS to assess the suitability of targets based on morphological features. The final velocity sample will extend into the North ($\delta\lesssim+10\deg$), increasing the area of the survey from $17,000$\,deg$^2$ to $20,600$\,deg$^2$. It will also increase the target density from $2$ to $2.5\ \mathrm{deg}^{-2}$ over the whole area. The Taipan Final peculiar velocity sample is therefore predicted to comprise up to $50,000$ galaxies over $2\pi$ steradians. This will include the brightest FP galaxies in the local universe ($z<0.1$) and will have uniform minimum spectral quality (S/N$\gtrsim 15\,$\AA$^{-1}$), which we expect will yield velocity dispersion measurement errors less than $10\%$. In Fig.~\ref{pv_nz}, we show the predicted redshift distribution for the Taipan Final peculiar velocity sample.

Based on our survey simulations (Section~\ref{observing}), which include a planned upgrade of the TAIPAN facility to 300 Starbugs available from the start of 2019, as well as reasonable assumptions about weather losses and instrument throughput, we expect our baseline Taipan Final survey to be completed 4.5 years after the start of survey operations (although this is dependent on the instrument meeting its performance specifications).

In the following sections, we describe our automated method for scheduling, acquiring, processing and archiving Taipan observations.

\subsection{Automated scheduling and fibre allocation}
\label{observing}

At the beginning of each night, the `virtual observer' software ({\em Jeeves}) generates an observing plan, including which fields to visit on that night, and which targets within each field to observe.
Optimal survey scheduling is akin to a travelling salesman problem, in which many of the different cities are seasonally and/or randomly inaccessible (according to weather).
Our strategy for solving this problem is to use `greedy' optimisation strategies (e.g.\,\citealt{Robotham2010}), which seek to choose the best available option at each step of the process.
In other words, rather than optimise over the entire life of the survey, the virtual observer identifies the best possible set of targets to observe with each successive pointing of the telescope.

\begin{figure*}
\centering
\begin{minipage}{\textwidth}
\includegraphics[trim={0cm 0cm 5cm 0cm},width=0.74\textwidth]{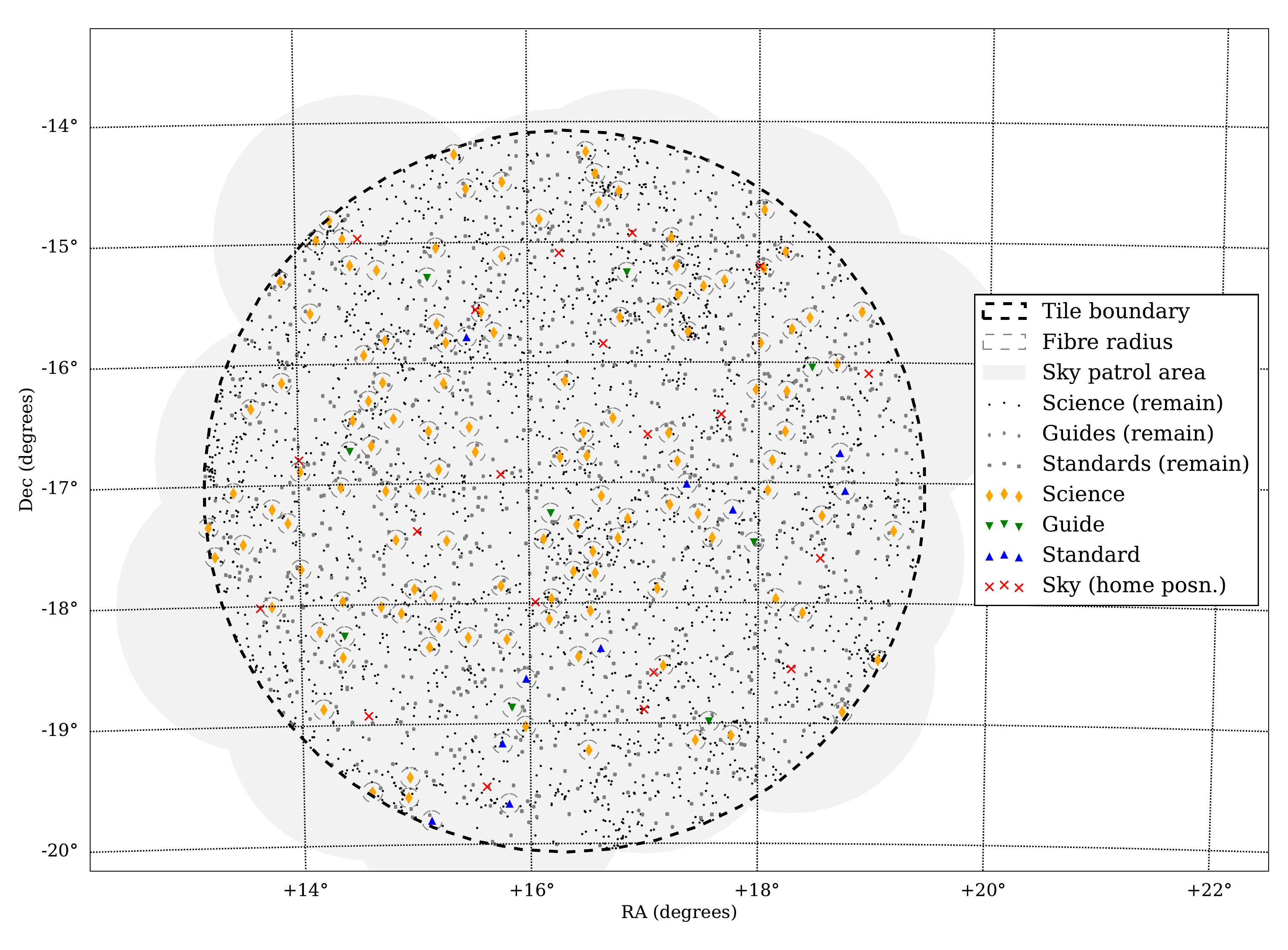}
\includegraphics[trim={0cm 0cm 5cm 0cm},width=0.74\textwidth]{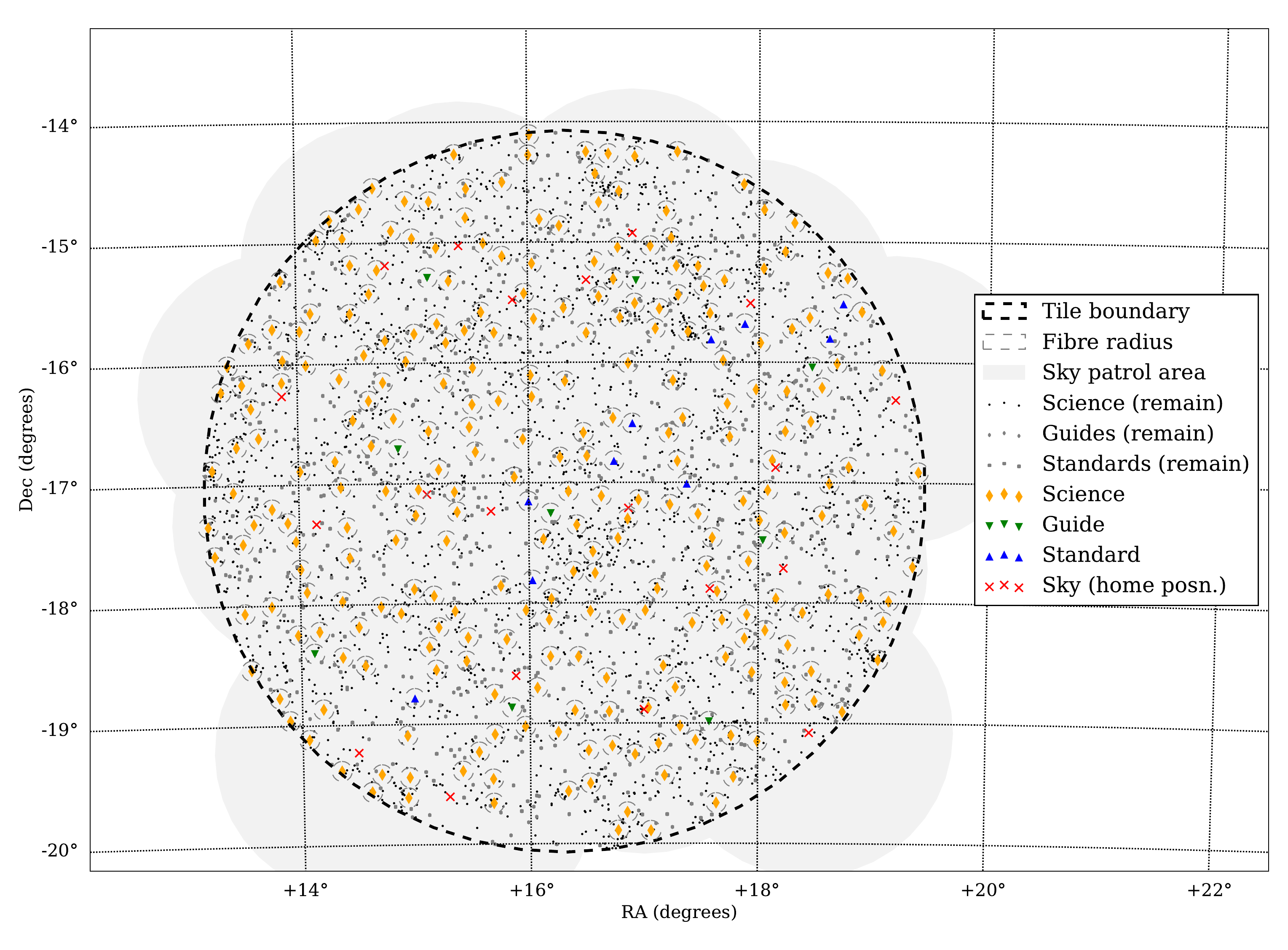}
\end{minipage}
 \caption{Example of fibre allocation by our tiling code in a single tile in the sky, using 150 Starbugs (top) and 300 Starbugs (bottom).} 
 \label{tiles}
\end{figure*}

The decision regarding which targets to observe next is a two-stage process. In the first stage, the best observable set of targets is determined for each one of a pre-defined set. The allowed pointings are defined using an optimal set of spherical coverings\footnote{See \url{http://neilsloane.com/coverings/index.html}} of possible pointings. For each potential pointing, fibres are allocated first to the highest priority targets. We continue to develop and refine the survey logic that will be used to determine the precise priority scores given to individual targets, but it is worth noting that target priorities will be reevaluated after each observation. For example, if a target is newly identified as a low-redshift early type galaxy (i.e., satisfying the selection criteria for the peculiar velocity sample), then it may become a high-priority target for repeat observation, to obtain the requisite S/N for a precise velocity dispersion measurement. For BAO science, where we prefer a high number of redshifts rather than completeness, an unobserved target has a higher priority than an already-observed one without a successful redshift determination. In this case, the target priority drops according to the number of times it has already been observed. To maximise our completeness in dense regions, where there are many targets of the same priority, preference is given to targets with the highest number of neighbours within the $10\,$arcmin fibre exclusion radius. In Fig.~\ref{tiles}, we show two examples of optimal Starbug allocation within a tile performed by our tiling algorithm, using 150 Starbugs (top panel), and 300 Starbugs (bottom panel).

Once the best possible tile (i.e.\,the set of targets with the highest net priority score) has been identified at all allowed pointings, the second stage is to select the best possible pointing to observe at the current time. Here, we devise a scalar figure of merit that provides an operational definition for the word `best', given the current sidereal time and state of the survey, and can be written as:
\begin{equation}
f = { P_\mathrm{allocated} \times N_\mathrm{remain}
		\over T_\mathrm{better} }.
\label{score}
\end{equation}
We define $P_\mathrm{allocated}$ as the summed priorities of all science targets allocated within a field.  
This value is modulated first by $N_\mathrm{remain}$, which is the number of high-priority targets in the field that have not yet been observed.  
This factor acts as a proxy estimate (up to some multiplicative scaling) for the expected number of times this field will need to be revisited to complete the survey.
We define $T_\mathrm{better}$ as the length of time between the time of observation and the anticipated end of the survey when the field is observable at its current airmass or less. 
$T_\mathrm{better}$ acts as a proxy estimate (up to some multiplicative scaling) for the expected number of opportunities to target this field that are as good or better than the present time.
To the extent that $N_\mathrm{remain}/T_\mathrm{better}$ represents the ratio between the number of times that a field needs to be revisited and the number of opportunities for a field to be revisited, it can be thought of as a characterisation of the `observing pressure' at that location on the sky. Similarly, the scheduling algorithm seeks to reduce the observing pressure by working towards the situation where $N_\mathrm{remain}/T_\mathrm{better}$ is uniform across the sky, and in so doing to minimise the amount of time needed to complete the survey.

The virtual observer will go through this process at the beginning of each night to generate that night's observing plan.  
Once {\em Jeeves} adds a field to the observing plan, all of the allocated targets within that field are temporarily removed from the target pool for subsequent fields, so that targets are not observed multiple times in the course of a single night.
At the end of each night, once all the data are taken and reduced, and the various quality control metrics have been evaluated, then each target is reevaluated and returned to the target pool if additional observations are required (for example, to achieve the desired S/N).
We note that while there is scope for more immediate feedback through the course of a night's observing, our intention at the beginning of survey operations is to go through the process of allocating and updating targets at most daily.

\begin{figure*}
\centering
\begin{minipage}{\linewidth}
 \includegraphics[trim={0.5cm 10cm 0.5cm 9cm},width=\textwidth]{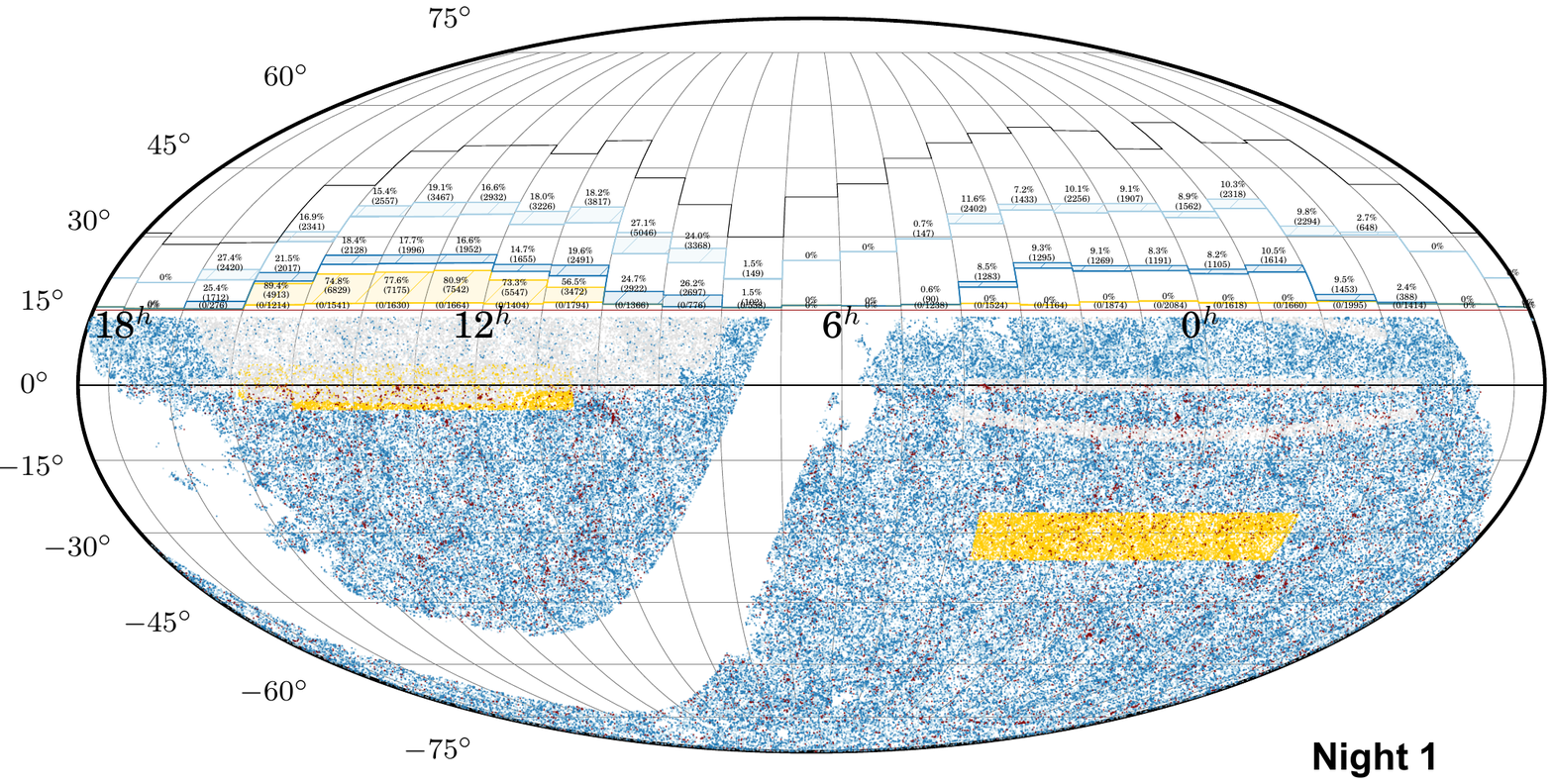}
  \includegraphics[trim={0.5cm 10cm 0.5cm 9cm},width=\textwidth]{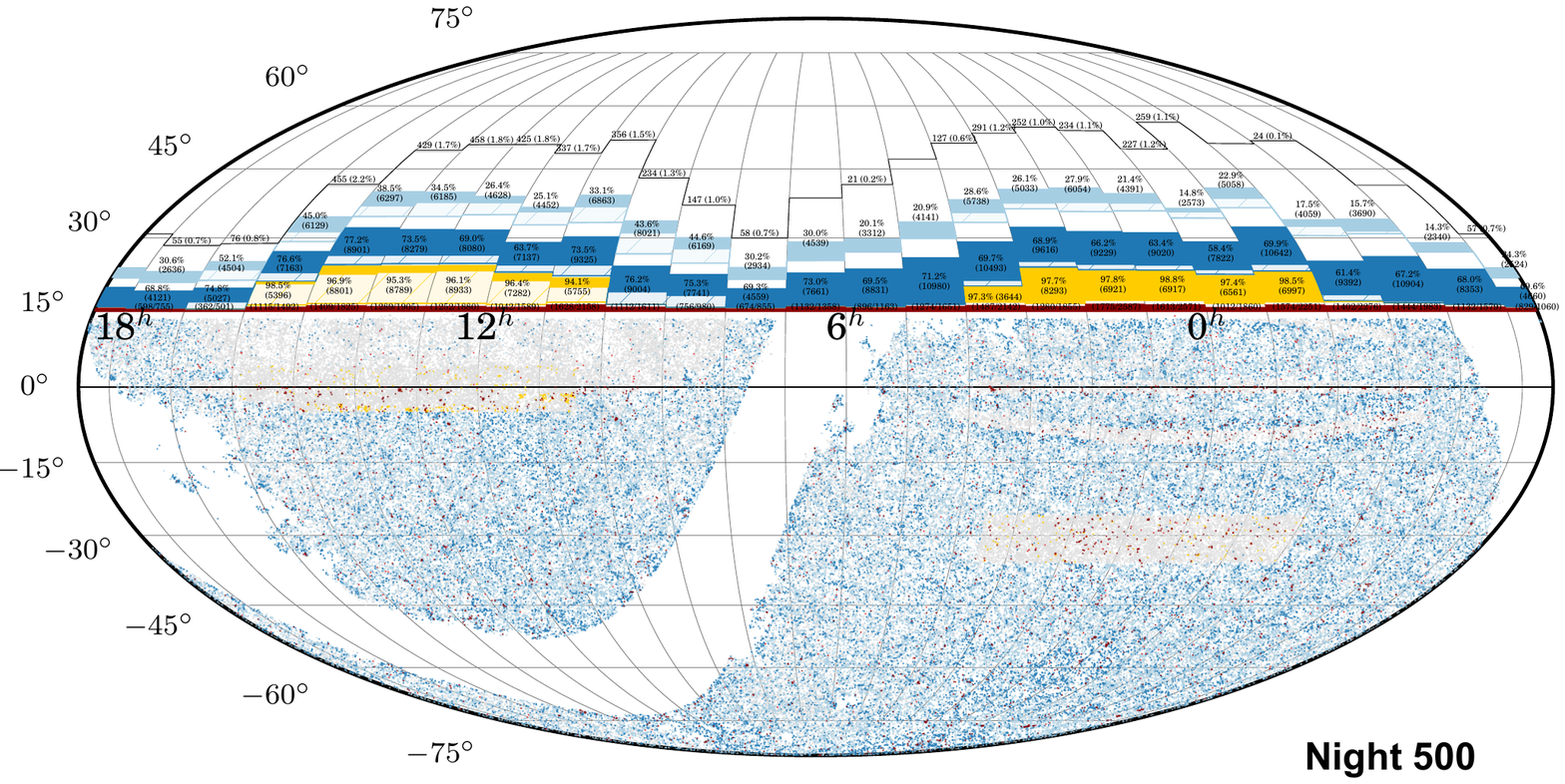}
  \end{minipage}
  \vspace{0.5cm}
 \caption{Mock sky distribution of our targets, obtained by applying our selection criteria to the {\sc galform} mock galaxy catalogue described in Section~\ref{galaxies}, and ensuring that we obtain the correct sky density by comparing to the GAMA survey over a smaller area. The top panel shows the output of our survey simulator at the start of observations (`night 1', set on September 1st, 2017 for this simulation), and the bottom panel shows the output after the 500th night of Taipan Phase~1 observations. Unobserved targets are colour-coded according to which subsample (Section~\ref{phase1}) they belong to: red -- peculiar velocity targets (dark red: 6dF-selected; light red: new early-type targets identified from the redshift survey); yellow -- complete magnitude-limited ($i\le17$) sample in the KiDS fields for galaxy evolution science; light blue -- magnitude-limited ($J_\mathrm{Vega}$<15.4) BAO targets; dark blue -- 2MASS-selected LRGs. The top histograms show the target RA distribution and the progress made within each sub-sample. As the survey progresses, completed targets turn light grey and the histograms become filled. The light grey targets and hashed histograms at the start of the survey represent SDSS sources for which spectra are already available -- we do not prioritise re-observing these targets which is why the SDSS footprint can be seen in the sky distribution.}
 \label{sim}
\end{figure*}

\begin{figure*}
\centering
  \includegraphics[trim={0cm 0cm 0cm 0cm},width=\textwidth]{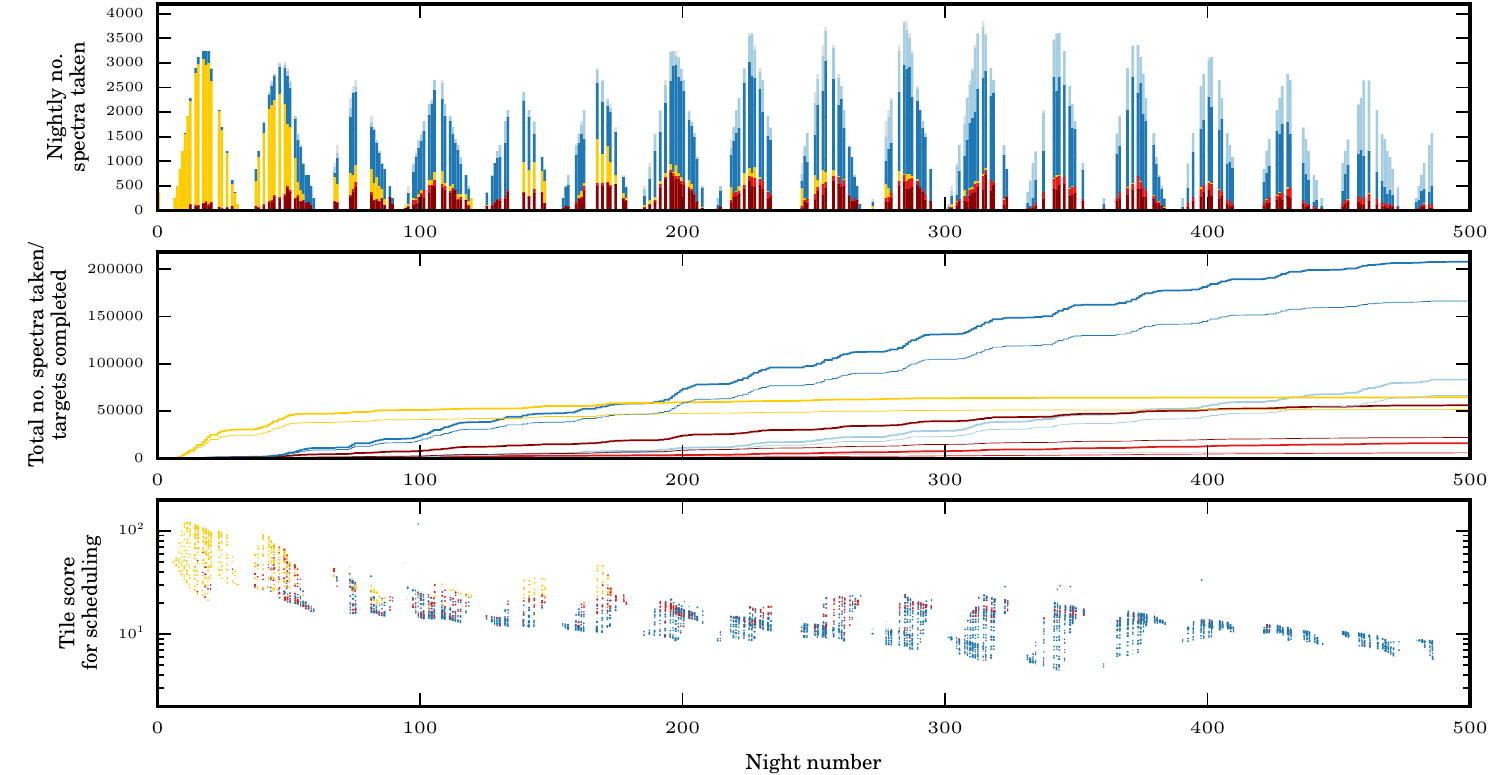}
 \caption{Simulated survey progress in Taipan Phase~1 predicted by our survey simulator (Section~\ref{observing}). The different colours refer to different sub-samples as described in Section~\ref{phase1}:  red -- peculiar velocity targets (dark red: 6dF-selected; light red: new early-type targets identified from the redshift survey); yellow -- complete magnitude-limited ($i\le17$) sample in the KiDS fields for galaxy evolution science; light blue -- magnitude-limited ($J_\mathrm{Vega}$<15.4) BAO targets; dark blue -- 2MASS-selected LRGs. {\em Top panel:} Number of spectra taken per night. The gaps correspond to random losses due to bad weather, and bright time. {\em Middle panel}: Cumulative number of spectra taken (including repeat observations of the same target; thick lines) and targets completed (i.e. redshift success and/or required S/N achieved; thin lines) in each subsample. {\em Bottom panel:} Tile scores ($f$; computing according to eq.~\ref{score}) in each night. Each tile is colour-coded according to the sub-sample from which the majority of its science targets comes from. The survey scheduling prioritises the spectroscopic completion of the magnitude-limited sample in the KiDS fields at the start of the survey when they are overhead, and then moves on to targeting BAO and peculiar velocity sources quasi-uniformly across the hemisphere, giving higher priorities to higher density regions (Fig.~\ref{sim}).}
 \label{simtime}
\end{figure*}

This approach to scheduling has been developed and validated through detailed simulations of our baseline survey, which combines our tiling/scheduling algorithms with some simple assumptions about observing cadence, weather loses, redshift success rates, etc. Specifically, we assume:
\begin{itemize}[topsep=0pt]
\item All dark time is
allocated for Taipan observing.
\item Probabilistic losses of $40\%$ of nights due to, e.g.\ bad weather. For simplicity, nights are considered lost in their entirety, without night-to-night correlations, seasonal variability, etc.
\item An observing cadence of 21 min 50 sec, which includes the $3\times5\,$min science integrations, 30 sec for an arc frame, plus $4\times20\,$sec for CCD readout, and $5\,$min for Starbug reconfiguration and telescope slew.
\item Probabilistic redshift success rate of $85\%$ per observation,
which amounts to a mean number of visits of $\sim1.4$ per object, in
accordance with simple expectations based on redshift success rates as a
function of S/N from previous survey experience (including 6dFGS and GAMA),
together with the SDSS fibre magnitude distribution of our targets, and the
anticipated TAIPAN instrument performance (Section~\ref{instrument}).
\item Probabilistic velocity dispersion success per-visit rates,
which amount to a mean number of visits of $\sim2.4$ per peculiar velocity
target, based on the requirement of S/N$\gtrsim15$, based on the SDSS fibre magnitude distributions of the 6dF-selected peculiar velocity targets.
\item Sample definitions and target distributions drawn from an
all-sky mock catalogue, based on the simulations described in Section~\ref{galaxies},
with care taken to match the expected all-sky target numbers/mean densities. We also mimic the availability of literature redshifts/spectra from SDSS.
\item The survey simulation begins on Sept 1, 2017, and runs for 16 months, i.e., up to the anticipated upgrade to 300 Starbugs.
\end{itemize}

The all-sky distribution of (mock) survey targets is shown in the upper panel of Fig.~\ref{sim}, with individual targets colour-coded according to how they are selected: $i$-band selected targets in the two KiDS fields in yellow; peculiar velocity targets in red; and 2MASS selected galaxies in blue. The footprint of the southern SDSS, where many literature redshifts and spectra are available, is clearly visible near the equator. The histograms around the top of this figure show the RA distributions of Taipan targets, using the same colour coding (the hatched regions show targets with redshifts and spectra available from SDSS).

In the lower panel of Fig.~\ref{sim}, we show the distribution of remaining targets after 16 months of Phase~1 survey operations. The $i$-band selected samples in the KiDS regions are almost entirely completed: nearly all of the yellow points are now grey. Our simulation obtains over $99.5\%$ targeting completeness, and $\approx98\%$ redshift completeness across KiDS-South\footnote{The completeness across KiDS-North is slightly less, owing to the difficulty of efficiently tiling/scheduling around the existing SDSS footprint. We continue to develop our strategy to improve this.}. Essentially all of the remaining peculiar velocity targets (red points in Fig.~\ref{sim}) are in dense clusters; this shows how well our sample will trace the large-scale structure and dark matter and baryonic mass distribution in the local Universe. We obtain velocity dispersion measurements for $\sim22,000$ ($\approx 70\%$) of our 6dF-selected targets, plus a further $\sim6,000$ new peculiar velocity targets identified from their Taipan spectra. We also observe $\sim210,000$ ($\approx80\%$) of $(J-K)$-selected LRG targets, which when combined with literature redshifts from 2dFGRS, 6dFGS, and SDSS should yield us close to $280,000$ redshifts for this LRG sample. Finally, we note the smooth and flat the distribution of remaining 2MASS-selected targets. which is by design, since the scheduling algorithm works to reduce the observing pressure across the sky.

In Fig.~\ref{simtime}, we show how the simulated survey progresses over the first 16 months. Observations of the $i$-selected targets in the KiDS-South field (yellow) are basically complete within the first 2--3 lunations, after which the scheduler shifts to prioritise all-sky LRG targets (dark blue). The southern edge of the KiDS-North field (where there is no SDSS coverage) is completed in lunations 4 and 5. After the first 12 months, the proportion of fibres allocated to peculiar velocity targets (red) begins to decrease because our ability to observe targets in this sample becomes limited by fibre collisions. The growing number of lower-priority 2MASS $J$-selected targets (light blue) after the first year shows where observability considerations push the scheduler towards prioritising greater completeness at extreme northern/southern declinations and/or relatively over dense fields, rather than an inability to efficiently tile the higher-priority $(J-K)$-selected LRG targets.

We continue to use our simulations to validate and optimise our observing strategy to ensure that we will achieve our ambitious goals. Nevertheless, these results already demonstrate our ability to efficiently obtain very high spectroscopic completeness across a very wide area. The most important current source of uncertainty is the actual performance of the TAIPAN instrument, especially regarding the exact Starbug reconfiguration time and the spectrograph throughput. These aspects will be precisely quantified in science commissioning operations before the start of the survey.

\subsection{Data processing strategy}
\label{data_processing}

\begin{figure*}
\centering
 \includegraphics[trim={0cm 9cm 0cm 0cm},width=\textwidth]{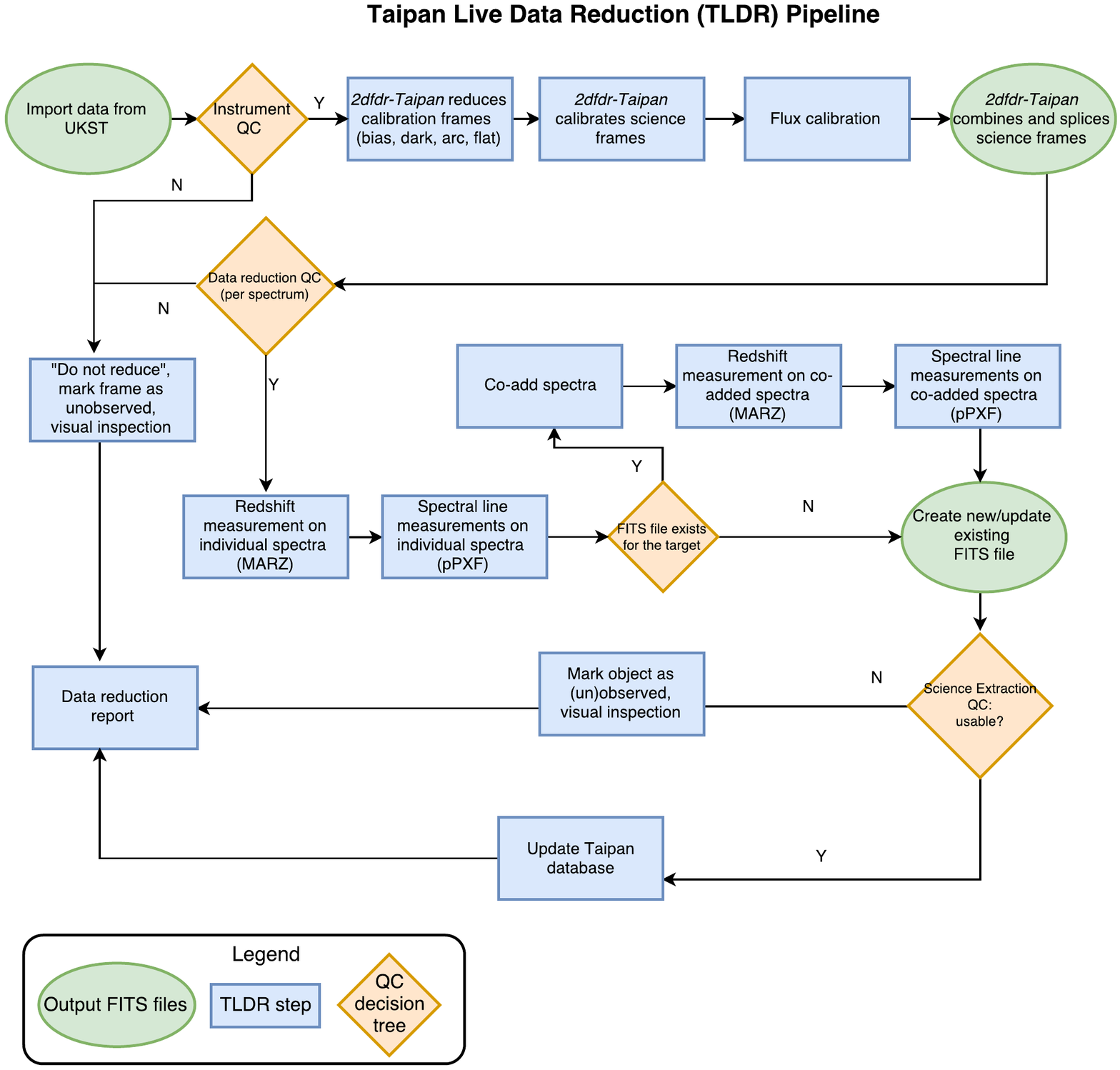}
 \caption{Flowchart describing our data processing strategy using the custom Taipan Live Data Reduction (TLDR) pipeline described in Section~\ref{data_processing}.}
 \label{tldr}
\end{figure*}

The Taipan galaxy sample data will be processed through our custom Taipan Live Data Reduction (TLDR) pipeline (Figure~\ref{tldr}).
As for the observing strategy, the goal is to have a fully-automated, machine-operated process from acquiring the data at the telescope, to
performing quality control tests, and producing calibrated, science-ready spectra and data products including redshifts, velocity dispersions, emission line fluxes, etc.
Here we briefly describe the four main steps of TLDR from raw data to science-ready products.

\subsubsection{Data reduction: 2dfdr-Taipan}

The first stage of TLDR uses a customised version of the {\em 2dfdr} 
multi-fibre spectroscopic data pipeline  \citep{2dfdr}, originally developed
in the mid-1990s for the 2-degree Field Galaxy Redshift Survey \citep{Colless2001} and
its spectrograph. {\em 2dfdr} has since been upgraded and updated to implement new surveys with other spectrographs (e.g.\,\citealt{Jones2004,Croom2012}).
For the Taipan survey, {\em 2dfdr} has been modified to accommodate TAIPAN's new spectral format.

The main task for {\em 2dfdr} in TLDR is to reduce the raw data by removing CCD artefacts,
and extracting individual spectra. This includes:
\begin{itemize}[topsep=0pt]
\item reducing bias and dark frames to obtain offsets in the spectra flux
levels caused by CCD noise;
\item reducing flat frames to perform tramline mapping to identify each spectrum
and derive fibre response curves along the wavelength direction;
\item reducing arc frames to identify key emission line positions and calibrate pixel coordinates to wavelength coordinates; and
\item using results from above to extract and wavelength-calibrate object spectra as well as removing
cosmic rays in each exposure.
\end{itemize}
Once these extracted spectra are obtained, they are ready for the following steps in the analysis: flux calibration, redshift determination and spectral measurements, plus quality control.

\subsubsection{Flux calibration}

Spectral flux calibration of Taipan is performed using F stars selected from the SkyMapper survey, 
following the approaches used by the SDSS and GAMA surveys.
We transform the SDSS broad-band colours to SkyMapper
broad-band colours using the colour terms measured by the SkyMapper team.\footnote{\url{http://skymapper.anu.edu.au/filter-transformations/}}
A SkyMapper colour cut is used to select F stars for flux calibration\footnote{\url{http://www.sdss.org/dr12/algorithms/boss_std_ts/}}:
\begin{multline}
\Big\{\big[(g-r)-0.18\big]^2+\big[(r-i)-0.11\big]^2 \\
+\big[(i-z)-0.01\big]^2\Big\}^{1/2} < 0.08.
\end{multline}
We note that we do not use the $u-g$ colour cut in the SDSS algorithm due to the red leak of the SkyMapper $u-$band filter.
We have tested our selection procedure using  SDSS spectroscopic observations, and find that about $80\%$ of the stars selected by these criteria are F stars (with the remainder are mostly G stars).

Spectral calibration stars from this list will be added into, and observed in, each Taipan field.
To flux-calibrate the observed galaxy spectra, we first restrict our photometrically-selected standard stars to those brighter than $r=16.5$ and with {\em a posterori}  acceptable spectral signal-to-noise. Based on the broad-band photometry, we then select and warp a synthetic spectral template from \cite{Pickles1998} to match the standard, before correcting for atmospheric extinction using the extinction coefficients measured at Siding Spring Observatory. A sensitivity function is then derived from a low-order spline fit to the ratio of the observed and warped synthetic spectra of the standard stars. Finally, the blue and red arm sensitivity curves are pieced together, and their spectra co-added. 

\subsubsection{Redshifts: \textsc{Marz}}

Having performed flux calibration and co-addition, all new and updated spectra are immediately redshifted. We automatically measure redshifts using \textsc{Marz} \citep{Hinton2016}, which implements a template-matching cross-correlation algorithm adapted from \textsc{Autoz} \citep{Baldry2014}. \textsc{Marz} fits input spectra against a range of stellar and galactic templates, and returns the redshift and template corresponding to the best cross-correlation, along with an estimate of the reliability (confidence level) of the result. \textsc{Marz} also allows easy visualisation of spectra via its web interface, however the primary usage of the application in our pipeline is to be run automatically without human input. \textsc{Marz} leverages a job queuing system, allowing fast redshift measurement and the potential to re-redshift prior targets in bulk if the data reduction pipeline undergoes improvement during the survey. The output redshifts and confidences from \textsc{Marz} are fed back into the survey database, where the optimal tile configurations and observational schedule for the telescope are updated.

\subsubsection{Spectral measurements and quality control}

After redshifts are determined, the next step is to perform further spectral measurements, using a custom version of the Penalised Pixel Fitting code (pPXF; \citealt{Cappellari2004,Cappellari2012}). We first mask known strong emission lines, and use pPXF to find the best-fitting simple stellar population (SSP) template combination, as well as an initial guess for the velocity dispersion and velocity offset (from the \textsc{Marz} redshift). We then rescale the {\em 2dfdr} variance array by the ratio of the standard deviation of the residuals after subtracting the best-fit templates. The next step is to unmask emission lines and include emission templates in the pPXF fit, as well as doing iterative cleaning to remove outliers before re-fitting (good variance estimates from the previous steps are needed for this clipping). This determines final estimates for the mean stellar velocity and velocity dispersion.

We then fix the stellar kinematics derived above and re-fit to determine the optimal combination of SSP templates for the underlying stellar continuum, again using pPXF and including emission templates. We interpolate the best-fit description of the stellar continuum onto the wavelength grid for the data and subtract from the data, leaving only the emission line residual spectrum. This step minimises the impact of re-binning multiple times on the emission line measurements and uncertainties determined in the next step, which consists of fitting Gaussians to emission lines in the residual spectrum ([OII] doublet, H$\delta$, H$\gamma$, H$\beta$, [OIII] doublet, [OI] doublet, [NII] doublet, H$\alpha$, [SII] doublet). The kinematics for the Balmer lines are tied together, as are, separately, the kinematics for the forbidden lines. The Gaussian amplitudes and widths are used to determine the line fluxes, and formal uncertainties are propagated through to determine a flux uncertainty. We also include a S/N proxy, which uses the standard formalism of \cite{Lenz1992} to estimate the line S/N based on the fit residuals.

The output spectral measurements and S/N are fed to the database, and the survey scheduler decides whether a target needs to be re-observed based on the survey rules, which prescribe a minimum required S/N for our targets.

\subsection{Data archiving and dissemination}
\label{archiving}

The Taipan galaxy survey data will be archived and made available to the wider community through a public database hosted by AAO Data Central, a node of the All-Sky Virtual Observatory (ASVO\footnote{\url{www.asvo.org.au}}). The Taipan database will include the final data products (reduced and calibrated spectra) and value-added catalogues including redshifts, spectral measurements, and multi-wavelength photometry, and will be accessible through a variety of mechanisms, including a web portal, simple Astronomical Data Query Language (ADQL) queries, and an application programming interface (API).

\subsection{Priority and ancillary science}
\label{ancillary}

Taipan's `parallel' science programmes sit in stark contrast to earlier multi-object surveys (like 6dFGS)
which used `spare fibres' not otherwise usable by the main science programme. Because of Taipan's relatively low fibre number compared to target number density, it will be able to allocate main targets to all fibres until essentially the very end of survey operations. As a result, Taipan ancillary science arising from additional targets that are not part of the main survey comes at a direct cost to main survey operations. We distinguish this from `priority science' that can be achieved through prioritising main survey targets or fields for early observation, which adds very little cost to the overall survey. 

Taipan's Priority and Ancillary Science programme recognises the value of enabling a broader range of science and providing opportunities for a wider community to participate in the Taipan program. Within our existing survey framework, there are three main avenues we are using to enable complementary science projects with Taipan spectroscopy:
\begin{enumerate}[topsep=0pt]
\item Prioritisation of some field(s) or target(s) for early completion (Priority Science); 
\item Repeat observation of some targets for monitoring or improved signal-to-noise ratio (Ancillary Science); 
\item Expansion of the Taipan sample to include additional targets/samples (Ancillary Science).
\end{enumerate}
The first option (i.e., priority targeting of particular fields or targets that are within the nominal Taipan sample) comes at very little additional cost to the main survey. Because Taipan will occupy essentially all of the dark time on the UKST for the duration of the survey, we have a degree of freedom and flexibility in choosing how to schedule or prioritise which targets/fields to observe. These flexible scheduling structures allow us to specify targeting priorities on the basis of position, observable properties (e.g. colour, brightness), or derived properties (e.g. redshift, line flux, equivalent width). To the extent that it is possible to schedule these fields/targets, adjusting the priorities does not have a large impact on the final content of the survey, or on the survey duration. The second and third options (i.e.\,expanding the sample to include additional targets, or spending extra time on selected main survey targets) are less straightforward, and come at additional operational cost, to be borne by the ancillary science proposers.

At the time of writing an initial set of priority and ancillary science programmes has been identified for the start of main survey operations. It is expected that there will be opportunities for further priority and ancillary programmes to be proposed as the survey progresses.

\section{Summary \& Conclusion}
\label{conclusion}

The Taipan galaxy survey will be conducted on the newly refurbished 1.2-metre UKST at Siding Spring Observatory  using the new AAO Starbug technology combined with a purpose-built spectrograph. It will carry out the most comprehensive spectroscopic survey of the southern sky to date, enabling high-precision measurements of cosmological parameters, as well as a new demographic study of the galaxy population in the local Universe.

In this paper we described the survey strategy, which is designed to optimally achieve three main goals:
\begin{enumerate}[label=(\roman*),topsep=0pt]
\item Measure the distance-scale of the Universe (principally governed by \ho) to $1\%$ precision using the baryon acoustic oscillations in the galaxy clustering pattern as a standard ruler. This will allow us to address current tensions between CMB and distance ladder measurements. It will also measure the growth rate of structure to $5\%$, which will allow us to test models of gravity.
\item Make the most extensive map constructed to date of motions in the local Universe using peculiar velocities, with a sample more than five times larger than available to 6dFGS. Combined with improved Fundamental Plane measurements, this will allow us to perform sensitive tests of the gravitational physics generating these motions.
\item Understand the baryon lifecycle and the role of mass and environment in the evolution of the galaxies, using spectroscopically-complete measurements, combined with HI measurements from the WALLABY survey.
\end{enumerate}
To achieve these scientific goals, Taipan will obtain optical spectra (from $370$ to $870\,$nm) for a magnitude-limited ($i\le17$, i.e.\, comparable to SDSS) sample of galaxies with near total completeness across the whole southern sky. This will be supplemented by a `luminous red galaxy' sample (selected to have $17<i<18.1$ and $g-i>1.6$) extending the survey volume as required for high-precision BAO measurements.

 Taipan will obtain about two million spectra over the whole southern hemisphere in 4.5 years. This survey speed and efficiency are enabled by the short field reconfiguration time enabled by the new Starbug technology. Taipan will be carried out in a fully automated way, and we have developed innovative software to optimally allocate targets from our input catalogues to each of the spectroscopic fibres in each 6-degree UKST field, to carry out each night's observing using the {\em Jeeves} virtual observer, and to process data through a data reduction and spectral measurement pipeline (TLDR).
The final Taipan database will include the final data products (reduced and calibrated spectra) and value-added catalogues including redshifts, spectral measurements, and multi-wavelength photometry from ancillary surveys such as SkyMapper, VHS, and WISE.

The legacy of Taipan will be a redshift and optical spectroscopic reference for the southern sky that is unlikely to be superseded for at least the next decade.

\section*{Acknowledgements}
We thank the referee for comments that helped us improve the clarity of the paper. We gratefully acknowledge funding support from the Australian Research Council through grants FT150100079 (EdC), DP160102075 (MC, CB), LE140100052 (MC), LP130100286 (JM), FL099213 (CW), DE150100618 (CL), FT140100255 (MSO), FT140101270 (KB).
Parts of this research were conducted by the Australian Research Council Centre of Excellence for All-sky Astrophysics (CAASTRO), through project number CE110001020.
JRL acknowledges support from the Science and Technology Facilities Council (STFC; ST/P000541/1).
MB is supported by the Netherlands Organisation for Scientific Research, NWO, through grant number 614.001.451.
We thank Michael Goodwin for producing Fig.~\ref{taipan_field}, and Camila Pacifici and St\'ephane Charlot for supplying model fits to SDSS galaxy spectra to test our pipeline.


\bibliographystyle{pasa-mnras}
\bibliography{bib_taipan} 

\end{document}